\documentclass[superscriptaddress,onecolumn,showpacs,a4paper,
amssymb,amsmath,nobibnotes,aps,prd,
showkeys,
nofootinbib,notitlepage]{revtex4-1}
\usepackage{verbatim}
\usepackage[T1]{fontenc}
\usepackage[utf8]{inputenc}
\usepackage[american]{babel}
\usepackage{epsfig}
\usepackage{graphicx,subcaption,caption}
\usepackage{booktabs}
\usepackage{multirow}
\usepackage{dcolumn}
\usepackage{amsmath}
\usepackage{mathtools}
\usepackage{amsfonts}
\usepackage{amssymb}
\usepackage{epstopdf}
\usepackage{bm}
\usepackage{siunitx}
\usepackage{braket}
\usepackage{enumitem}
\usepackage{soul}
\usepackage[table]{xcolor}
\usepackage{color}
\usepackage{transparent}
\usepackage{pifont}

\definecolor{navyblue}{rgb}{0.0, 0.0, 0.5}
\definecolor{royalblue}{rgb}{0.25, 0.41, 0.88}
\definecolor{cadmiumgreen}{rgb}{0.0, 0.42, 0.24}
\definecolor{blue-violet}{rgb}{0.54, 0.17, 0.89}
\definecolor{darkviolet}{rgb}{0.58, 0.0, 0.83}
\definecolor{orange(colorwheel)}{rgb}{1.0, 0.5, 0.0}

\usepackage{hyperref}
\hypersetup{
    colorlinks=true, 
    linkcolor=royalblue, 
    citecolor=magenta}

\usepackage{booktabs}
\usepackage{multirow}
\usepackage{dcolumn}
\usepackage{colortbl}
\newcommand\vertsp{\rule[-2mm]{1mm}{0mm} &}
\newcommand\horsp{\rule[-1.5mm]{0mm}{4.125mm}}
\newcommand\morehorsp{\rule[-2.25mm]{0mm}{6mm}}
\newcommand\ie{\emph{i.e.}}

\begin{document}

\title{Propagating Speed of Primordial Gravitational Waves.}

\author{William Giarè}
\email{william.giare@uniroma1.it}
\affiliation{Physics Department and INFN, Universit\`a di Roma ``La Sapienza'', Ple Aldo Moro 2, 00185, Rome, Italy}

\author{Fabrizio Renzi}
\email{renzi@lorentz.leidenuniv.nl}
\affiliation{%
Lorentz Institute for Theoretical Physics, Leiden University, PO Box 9506, Leiden 2300 RA, The Netherlands
}%

\date{\today}
\preprint{}
\begin{abstract}
Primordial Gravitational Waves, i.e. a background of metric perturbations sourced by the quantum inflationary fluctuations, if measured, could both provide a substantial evidence for primordial inflation and shed light on physics at extremely high energy scales. 
In this work we focus on their propagating speed. Using an effective field theory approach we introduce a time-dependent propagating speed $c_{\rm T}(t)$ showing that also small deviations from the General Relativity (GR) prediction $c_{\rm T}(t) = c$ can lead to testable consequences. We derive a set of equations that relate the propagating speed and its time dependence to the inflationary parameters and that generalize the usual slow roll consistency relations.
Imposing the new generalized consistency relations and combining small and large scales data,
we derive model independent constraints on inflation with non-trivial primordial tensor speed. In particular we constrain its scale dependence to be $d\log c_{\rm T} / d\log k=0.082^{+0.047}_{-0.11}$ at 68\% C.L. while we only derive the lower bound $c_{\rm T}>0.22\,c$ at 95\% C.L. .
We also constrain the tensor-to-scalar ratio at the pivot scale $k_*=0.05\rm{Mpc}^{-1}$ to be $r<0.0599$ at 95\% C.L. in agreement with the result provided by the Planck Collaboration. Thanks to a proper small scale parametrization of the tensor spectrum we derive stringent constraints on the tensor tilt $n_{\rm T}=-0.084^{+0.10}_{-0.047}$ at 68\% C.L. and on its runnings $\alpha_{\rm T}=d\,n_{\rm T}/d\log k=0.0141^{+0.0035}_{-0.021}$ and $\beta_{\rm T}=d\,\alpha_{\rm T}/d\log k= -0.0061^{+0.010}_{-0.0014}$  both at 68\% C.L.
Our results show a remarkable agreement with the standard slow roll predictions and prove that current data can significantly constrain deviations from GR on the inflationary energy scales.

\end{abstract}


\maketitle

\section{Introduction} \label{Section_1}
Primordial Inflation \cite{Guth:1980zm}, a phase of accelerated expansion of the early universe, cannot only solve all the Hot Big Bang Theory shortcomings, but it can also make some predictions.
In fact, the quantum inflationary fluctuations can both explain the scalar perturbations observed in the Universe and predict a background of metric perturbations, known as Primordial Gravitational Waves (PGWs) \cite{Linde:1981mu,Vilenkin:1983xq,Lyth:2009zz,Mukhanov:2005sc,Starobinsky:1980te,Weinberg:2008zzc,Martin:2013tda,Riotto:2018pcx}.  
The detection of B-modes in the Cosmic Microwave Background (CMB) polarization originated from the inflationary tensor modes and, in general, the detection of the PGWs is one of the most important goals of modern cosmology since they can both provide a substantial evidence for primordial inflation and shed light on its physical nature \cite{Baumann:2014cja,Caldwell:2018giq,Franciolini:2018ebs,Kamionkowski:2015yta}.
At least in the simplest models, the total amount of PGWs is proportional to the energy scale at which inflation occurs and a satiable background of PGWs is expected at sufficiently high energy scales \cite{Lyth:2009zz,Mukhanov:2005sc,Dodelson:2003ft,Weinberg:2008zzc,Martin:2013tda,Kamionkowski:2015yta,Mirbabayi:2014jqa,Mirbabayi:2014jqa,Ozsoy:2014sba,Riotto:2018pcx}. PGWs can therefore provide information about the theory of gravity at extremely high energy and consequently they can be used to test General Relativity (GR).  

In the recent years the bound on the amplitude of PGWs from CMB data, parametrized through the so-called tensor-to-scalar ratio $r$, has witnessed significant improvement. An upper limit $r_{0.002} < 0.056$ at $95\%$ C.L. has been provided in the last data release of the Planck Collaboration \cite{Akrami:2018odb} combining Planck and BICEP2/Keck array (BK15) data \cite{Ade:2018gkx}. An improvement of an order of magnitude with respect to the first constraints from the BICEP experiment of $r < 0.72$ at $95\%$ C.L. \cite{Chiang:2009xsa}.
In the upcoming decade, a new generation of CMB experiments (e.g. BICEP3 \cite{BICEP3}, CLASS \cite{CLASS} , SPT-3G \cite{SPT-3G}, Advanced ACTPol \cite{ACTPol}, LBIRD \cite{LBIRD} and CMB-S4 \cite{CMB-S4}) is expected to bring the sensitivity to the amplitude of tensor perturbations down to $r \sim 0.01 - 0.001$ improving the current Planck upper limit around an order of magnitude and possibly leading to the first detection of non-zero tensor amplitude. 
However these bounds on the tensor amplitude are derived assuming the usual consistency relation between the tensor spectral index $n_{\rm T}$ and the tensor-to-scalar ratio, namely  $n_{\rm T} = -r/8$, basically leading to an almost flat tensor spectrum  (see also \cite{Renzi:2019ewp,Shokri:2019rfi}). 
In practice the consistency relation between $r$ and $n_{\rm T}$ 
is  violated in many (non standard) models of inflation\footnote{For example one can consider a mechanism of sourced gravitational waves form a rolling spectator axion coupled with gauge fields during inflation \cite{Mukohyama:2014gba,Namba:2015gja,Peloso:2016gqs,Giare:2020vhn,Ozsoy:2020ccy}, or even more elaborated scenarios \cite{Stewart:2007fu,DEramo:2019tit}.} and in most of them it is no more possible to fix the energy scale of inflation from a direct measurement of the tensor modes amplitude. Moreover, when the inflationary consistency relation is relaxed (\emph{i.e.} $n_T \neq -r/8$), Planck data only weakly constrain the tensor tilt $n_{\rm T}$ to be $-0.55<n_{\rm T}<2.54$ \cite{Akrami:2018odb}. 
Combining CMB data with ground-based gravitational waves interferometers data, the upper bound on the tensor tilt is further improved to $n_T < 0.52$ \cite{Akrami:2018odb}.

In fact, along with B-modes polarization, primordial gravitational waves may also imprint the so-called stochastic gravitational waves background, the analogous of CMB for gravitational waves \cite{Caprini_2018}. 
While a direct detection of the stochastic background has not yet been provided, the first and second observing runs of the LIGO/VIRGO Collaboration placed an upper bound on its amplitude for the scales $k_{\rm LV} \in \left(1.3\,\rm{-}\,5.5\right),\times 10^{16} \,\rm{Mpc}^{-1}$ \emph{i.e.} 
\begin{equation}
\Omega_{\rm{GW}} (k_{\rm LV}) \leq 1.7 \times 10^{-7}.
\label{LVlimit}
\end{equation}
at 95\% C.L.\cite{LIGO_SGWB-2017,LIGO_SGWB-2019}. Assuming that the power law approximation for primordial spectra is valid from these ultra-high $k$ all the way up to the CMB scales \footnote{It is worth noting that the Planck Collaboration has shown that the assumption of a pure power law for the primordial spectrum is valid at least between the scales proved by Planck data \emph{i.e.} $0.005\, \rm Mpc^{-1} \lesssim k \lesssim 0.2\, \rm Mpc^{-1}  $ where primordial perturbations are linear. We also note that recently it has been shown that bounds on the stochastic background can be derived on CMB scales from CMB data alone assuming that gravitational waves behave as an effective neutrino species \cite{Clarke:2020bil} \emph{i.e} that PGWs effectively contribute to the total number of relativistic species at recombination \cite{Clarke:2020bil,Cabass:2015jwe}}, the LIGO/VIRGO constraint on the amplitude of the stochastic background can be translated into constraints on the primordial tensor modes \cite{Akrami:2018odb,Bartolo:2016ami,Cabass:2015jwe,Stewart:2007fu,Wang:2016tbj}.

In fact the fraction of the energy density of the universe due to PGWs at the present time and at a given scale $k=2\pi\,f$ is \cite{Akrami:2018odb,Bartolo:2016ami,Cabass:2015jwe,Stewart:2007fu}
\begin{equation}
\Omega_{\mathrm{GW}}(k) \doteq \frac{1}{\rho_{c}} \frac{\mathrm{d} \rho_{\mathrm{GW}}}{\mathrm{d} \log k}=\frac{\mathcal P_{\mathrm{T}} (k)} {24 z_{\mathrm{eq}}}
\end{equation}
where $\mathcal P_{	\rm T }$ is the primordial tensor spectrum at the scale $k$ and $z_{\rm{eq}} \sim 3400 $ is the redshift at the matter-radiation equivalence \cite{Akrami:2018odb}.

\begin{figure}[h]
    \centering
    \includegraphics[width=0.7\textwidth]{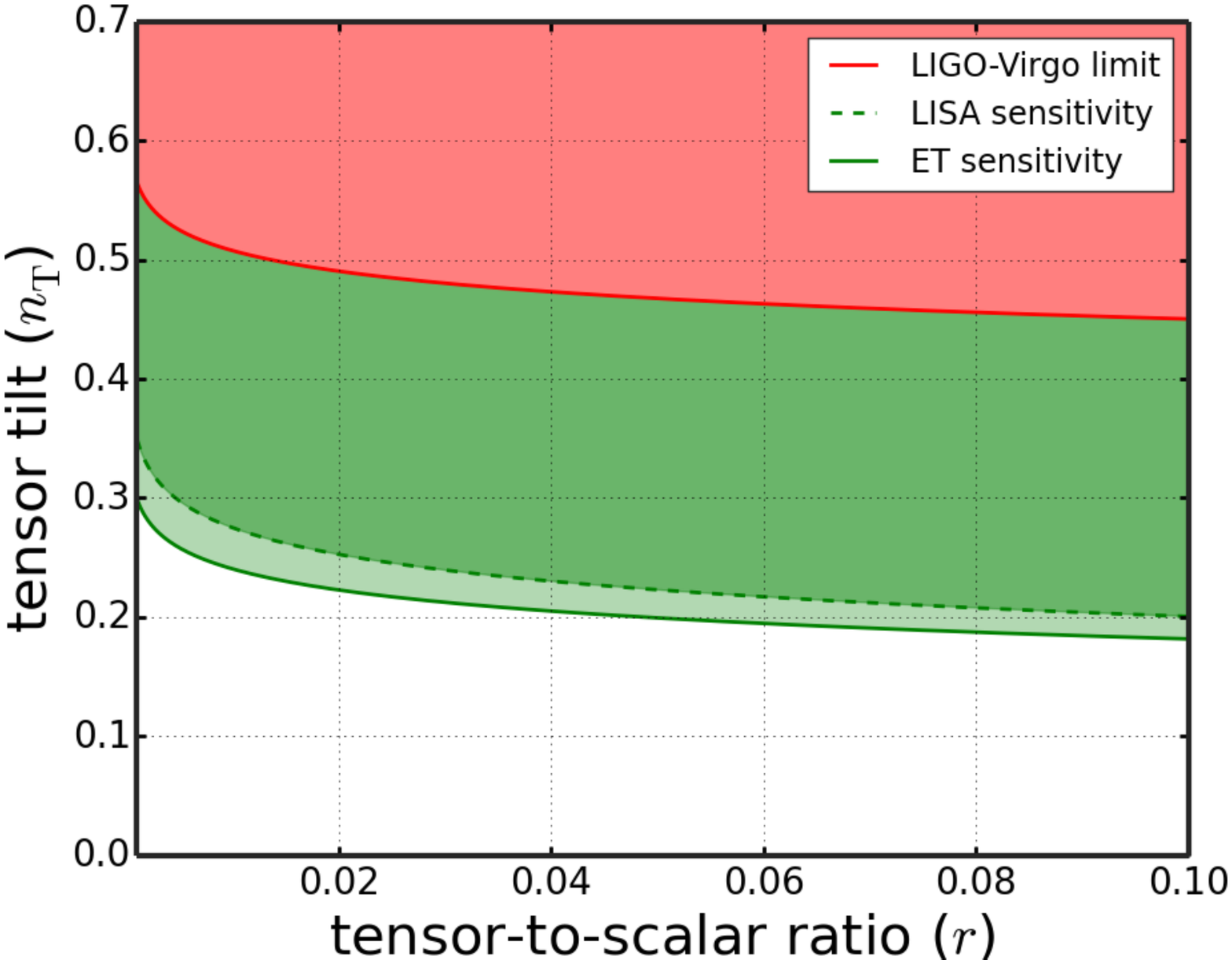}
    \caption{Small-scale constraints in the plane ($r$ , $n_{\rm T}$) derived 
    by $\Omega_{\rm GW}(k)$ trough Eq.\eqref{blue_tilt} or equivalently its scale-dependent generalization \eqref{Blue_tilt(k)}. The red region is excluded at 95\% C.L. by the LIGO/VIRGO limit \eqref{LVlimit} (red solid line). The green regions represent the sensitivity of future GWs experiment, such as LISA \cite{Audley:2017drz} (green dashed line) and ET \cite{Punturo:2010zz} (green solid line). If the tensor tilt is assumed to be scale-independent all these constraints refer also to its value on the CMB scale: $n_{\rm T}(k_*)$. Instead in the scale-dependent parametrization eq.\eqref{nT(k)} they refer to different scales: $n_{\rm T}(k_{ET})\lesssim n_{\rm T}(k_{\rm LISA}) < n_{\rm T}(k_{\rm LV})$, as discussed in Sec. \ref{sec.LVconstraints}.}
    \label{fig:figure1}
    \end{figure}

For a power-law tensor spectrum, taking $\mathcal{P}_T(k_*) = r\mathcal P_{\rm S}(k_*)$ where $\mathcal P_{\rm S}(k_*)$ is the amplitude of scalar perturbation at $k_{*} = 0.05\,\rm{Mpc}^{-1}$, one can translate an upper bound on the stochastic background $\Omega_{\rm GW}(k)$ into a constraint on the tensor tilt :
\begin{equation}
n_{\rm T} <\frac{\ln\left(\frac{24\,z_{\rm eq}\,\Omega_{\rm GW}(k)}{r\,\mathcal P_{\rm S}(k_*)}\right)}{\ln\left(\frac{k}{k_*}\right)} \lesssim 0.5
\label{blue_tilt}
\end{equation}
where in the last inequality we have evaluated the expression at $k=k_{\rm LV}$ using the LIGO/VIRGO limit \eqref{LVlimit} and taking $\mathcal P_{\rm S}(k_*) = 2.1\times 10^{-9}$ and $r \sim 10^{-2} $.  The next generation of gravitational waves probes (such as LISA \cite{Audley:2017drz} and Einstein Telescope \cite{Punturo:2010zz}) are expected to bring this upper limit down by a factor of $\sim 2$ (see Fig. \ref {fig:figure1}) \footnote{We assumed LISA to have a sensitivity to the stochastic background $\Omega_{\rm{GW}}(k_{\rm Lisa})\simeq 1\times10^{-12}$ on scales $k_{\rm Lisa}\approx 1\times10^{13} \rm{Mpc}^{-1}$ \cite{Bartolo:2016ami} while for the Einstein Telescope we assumed a sensitivity of $\Omega_{\rm{GW}}(k_{\rm ET})\simeq 3\times10^{-13}$ on scales $k_{\rm ET}\approx 5\times 10^{15} \rm{Mpc}^{-1}$  \cite{Maggiore:2019uih}}. If the tensor tilt is assumed to be scale-independent, these bounds clearly refer to the CMB scales $n_{\rm T}(k_*)\equiv n_{\rm T}(k),\ \forall k$.

However due to the huge difference in the scales proved by CMB and GW data, non-linearities may significantly affect the shape of the primordial spectrum possibly breaking the power-law assumption \cite{Chongchitnan:2006pe,Friedman:2006zt,Smith:2006xf}. When non-linear corrections are considered the higher-order terms in the primordial spectrum (i.e. the runnings \cite{Giare:2019snj,Kuroyanagi:2011iw,Zarei:2014bta}), even if tiny on CMB scales, may lead to non-negligible corrections on smaller scales where the amplitude of PGWs is proved by gravitational detectors and cannot be ignored when constraints are derived from such data. In fact high-order corrections may non-trivially connect the constraints on the CMB scales with the constraints on the astrophysical scales,  \ie  $n_{\rm T}(k)\neq n_{\rm T}(k_*)$, so that, depending on the model, an improvement in the constraints on astrophysical scales may or may not lead to an improvement in the constraints on the CMB scales.

The increased precision in the constraints on the primordial tensor modes from the current (and future) small and large scale experiments opens up the possibility of probing the physics of inflation with primordial gravitational waves, testing deviations from the standard slow roll predictions as a hint for new physics. It is therefore timely to investigate which constraints one can obtain from current CMB and GW data on inflationary models that can lead to deviations from the standard inflationary consistency relations.

In this paper we focus on models with a non-trivial propagating speed of primordial gravitational waves. In GR the propagating speed of the gravitational waves, $c_{\rm T}$, is the same as the speed of light $c$. Thus, working in the natural units, one can set $c_{\rm T }=c=\hslash=1$.  However, this cannot be true in more elaborated modified gravity theories such as the Horndeski theory of gravity \cite{Horndeski:1974wa,Deffayet:2011gz,Kobayashi:2011nu,Gao:2014soa,Gao:2014fra,Gleyzes:2014qga}, the Gauss-Bonnet gravity \cite{Nojiri:2005jg,Makarenko:2017vuk,Bamba:2014zoa,Feng:2020duo,Odintsov:2020xji,Oikonomou:2020sij, Odintsov:2020xji,Odintsov:2020zkl}, and also the low-energy effective string theory with higher-order corrections \cite{Satoh:2008ck,Baumann:2015xxa,Oikonomou:2015qha,Haro:2015oqa,Ballesteros:2014sxa,Antoniadis:1993jc,Kawai:1998ab,Soda:1998tr,Kawai:1999pw,Cartier:1999vk,Cartier:2001is,Piao:2003hh}. Even if the propagating speed of the astrophysics Gravitational Waves is measured from the ground-based interferometers and it is in good agreement with the speed of light \cite{Monitor:2017mdv, Cornish:2017jml,Liu:2020slm} (see also \cite{Bonilla:2019mbm} for forecasts at high  redshift), the propagating speed of the cosmological primordial gravitational waves, albeit previously discussed in literature \cite{Raveri:2014eea,Creminelli:2014wna,Giovannini:2015kfa,Cai:2015ipa,Cai:2015yza,Cai:2016ldn,Fumagalli:2016afy, Gao:2019liu,Noumi:2014zqa,Bordin:2017hal}, is currently essentially unconstrained. This is because the lack of a direct detection of the tensor spectrum makes it difficult to constrain the PGWs propagation with high precision. Nevertheless, any deviation from a constant $c_{\rm T}=1$ would imply new physics beyond GR, so constraining the propagating speed of PGWs and its time dependence means to test gravity literally at the earliest moments of time when inflation takes place and the primordial tensor modes are generated by the quantum inflationary fluctuations. 
Using an effective field theory approach, we introduce a generic time-dependent propagating speed $c_{\rm T}(t)$ and, under the assumption of slow roll inflation, we show that also small deviations from the GR condition $c_{\rm T} = 1$ can leave testable consequences in the inflationary parameters. In particular such models may lead to blue tilted tensor spectrum and affect the small scale behaviour of tensor anisotropies. In the following, we derive a set of equations that relate the propagating speed to the inflationary parameters and that generalize the usual slow roll consistency relations that are, in fact, recovered when the GR prescription $c_{\rm T}=1$ is restored. Imposing the new generalized consistency relations and combining the CMB data together with the small scales data on the stochastic background of Gravitational Waves, we are able to provide model independent constraints on the inflationary parameters.
The paper is organized as follows: in Sec. \ref{Theory} we review the theory of the tensor inflationary fluctuations allowing the possibility to have a non-trivial time dependent propagating speed $c_{\rm T}(t)$ during the inflation. In Sec. \ref{sec.generalcons} we investigate the consequences of the non-trivial propagating speed on the inflationary parameters. We derive a set of equations that generalize the usual slow roll consistency relations and that relate $c_{\rm T}(t)$ to the tensor spectral parameters. The modified consistency relations provide a powerful method to constrain the propagating speed and its time dependence allowing us to test gravity on the inflationary energy scales. In Sec. \ref{Constraints}, imposing the generalized consistency relations, we first derive some constraints using the most recent CMB data. Then we derive other constraints from small scale experiments on gravitational waves. Finally we combine the CMB and the small scale constraints improving the final results. In Sec. \ref{sec.conclusion} we present our conclusions.

\section{Theory}
\label{Theory}
In this section we briefly review the theory of the primordial tensor perturbations during inflation \cite{Riotto:2018pcx, Guth:1985ya,Mukhanov:1990me,Starobinsky:1979ty,Mukhanov:2013tua,Bartolo:2001rt, Weinberg:2008hq} introducing a non-trivial propagating speed $c_{\rm T}(t)$. We perform an approach based on the effective field theory of inflation (EFT): the action for the single field inflation in the unitary gauge is \cite{Cheung:2007st,Weinberg:2008hq,Burgess:2017ytm,Baumann:2014nda}:
\begin{equation}
S=\frac{M_{p}^{2}}{2} \int d^{4} x \sqrt{-g} \left[R-c_{1}(t)-c_{2}(t) g^{00} -\left(1-\frac{1}{c_{\rm T}^{2}(t)}\right)\left(\delta K_{\mu \nu} \delta K^{\mu \nu}-\delta K^{2}\right)\right]
\end{equation}
where $M_p^2=\frac{1}{8\pi\rm G}$, $c_1(t)=2\left(\dot H +3H^2\right)$, $c_2(t)=-2\dot H$ and $K_{\mu\nu}$ is the extrinsic curvature of the spatial slices. Here a dot denotes the derivative with respect to the cosmic time $\dot x \equiv dx/dt$. Note that in the standard slow roll case ($c_{\rm T}=1$) the part of the action involving the extrinsic curvature vanishes and we recover the standard action in the unitary gauge. Moreover a non-trivial propagating speed $c_{\rm T}(t)$ does not affect the spectrum of the scalar perturbation. Therefore we can consider only the tensor perturbations whose quadratic action reads
\begin{equation}
S_{\gamma}^{(2)}= \frac{M_{p}^{2} }{8}\int d \tau \, d^{3} x  \frac{a^2}{c_{\rm  T}^2(t) }\left[\left(\frac{d \gamma_{i j}}{d \tau}\right)^{2}-c_{\rm T}^{2}(t)\left(\vec{\nabla} \gamma_{i j}\right)^{2}\right]
\end{equation}
where $a(t)$ is the scale factor, $d\tau=dt/a(t)$ is the conformal time and $\gamma_{i j}$ is transverse and traceless: $\gamma_{ii}=0$ and $\partial_i \gamma_{ij}=0$. We expand $\gamma_{i j}$ in the Fourier series:
\begin{equation}
\gamma_{i j}(\tau, \mathbf{x})=\int \frac{d^{3} k}{(2 \pi)^{3}} e^{-i \mathbf{k} \cdot \mathbf{x}} \sum_{p=+, \times} \gamma_{p}(\tau, k) a_{(p)}(\mathbf{k}) \, \lambda_{i j}^{(p)}(\mathbf{k}) +\rm{h.c.}
\end{equation}
where the sum is over the polarization states $p=(+,\times)$ and the polarization tensor $\lambda_{ij}^{p}(\textbf{k})$ satisfies the usual conditions
\begin{align}
&k_j\lambda_{ij}^{(p)}(\textbf{k})=0, \\
&\lambda_{ii}^{(p)}(\textbf{k})=0 \\
& \lambda_{ij}^{(p)}(\textbf{k})\lambda_{ij}^{*\,(p')}(\textbf{k})=\delta_{pp'}\\
& \lambda_{ij}^{*\,(p)}(\textbf{k})= \lambda_{ij}^{*\,(p)}(-\textbf{k})
\end{align}
as well as the creation and annihilation operators satisfy 
\begin{equation}
\left[ a_{(p)}(\textbf{k}) \,,\, a^{\dagger}_{(p')}(\textbf{k}')\right]=\delta_{pp'}\,\delta^3(\textbf{k}-\textbf{k}').
\end{equation}
It is trivial to check that, defining the fields
\begin{equation}
u(\tau, k)\doteq\gamma_{(p)}(\tau, k) z_{T}, \quad z_{T}\doteq\frac{M_{p}}{2}\left(\frac{a}{c_{\rm T}(t)}\right)
\label{zT}
\end{equation} 
the equation of motion reads
\begin{equation}
\frac{d^{2} u}{d \tau^{2}}+\left( c_{T}^{2} k^{2}-\frac{1}{z_{T}} \frac{d^{2} z_{T}} {d \tau^{2}}\right) u=0.
\label{Eq_motion}
\end{equation}
In what follows we work under the following conditions. First of all we fix a background slow roll dynamics requiring that $|\dot H|\ll H^2$. So we define the slow roll parameters
\begin{equation}
\epsilon_1\doteq -\frac{\dot H}{H^2},
\end{equation} 
\begin{equation}
\epsilon_{i>1}\doteq\frac{d\log \epsilon_{i-1}}{d\log k}\simeq\frac{\dot \epsilon_i}{H\epsilon_i},
\end{equation} 
with $0<\epsilon_1\ll 1$ in such a way that the Null Energy Condition (NEC) is preserved and $|\epsilon_{i>1}|\ll1$. Moreover we also assume the variation of the propagating speed per Hubble time to be small, defining the similar parameters
\begin{equation}
\epsilon_1^{\rm T}\doteq \frac{\dot c_{\rm T}(t)}{H\,c_{\rm T}(t)} ,
\label{e1T}
\end{equation} 
\begin{equation}
\epsilon_{i>1}^{\rm T}\doteq \frac{d\log \epsilon^{\rm T}_{i-1}}{d\log k} \simeq \frac{\dot\epsilon^{\rm T}_{i-1}}{H\epsilon^{\rm T}_{i-1}},
\label{et2}
\end{equation}
with $|\epsilon_1^{\rm T}|\ll 1$ and $|\epsilon_{i>1}^{\rm T}|\ll 1$. In this way one can show that
\begin{equation}
\frac{1}{z_T}\frac{d^2 z_T}{d\tau^2}\simeq\frac{1}{a}\frac{d^2 a}{d\tau^2}\simeq\frac{2}{\tau^2}
\end{equation}
at least corrections of order $\epsilon$ (see Appendix \ref{App.A} for further details). Moreover, it is also easy to check that one can define a new wave vector $\tilde k \doteq c_{\rm T}(t) k$ that can be regarded as constant in the conformal time since its derivative is of order $\epsilon_1^{\rm T}$. At the end of the game, unless corrections of order $\epsilon$, we can write our equation as 
\begin{equation}
\frac{d^{2} u}{d \tau^{2}}+\left( \tilde k ^2-\frac{2}{\tau^2} \right) u=0
\label{Eq_motion_2}
\end{equation}
with the solution (obtained fixing the Bunch-Davies vacuum)
\begin{equation}
u(\tau,\tilde k)=\frac{e^{-i \tilde k \tau}}{\sqrt{2 \tilde k}}\left(1-\frac{i}{\tilde k \tau}\right).
\label{Solutuion}
\end{equation} 
A more detailed derivation of this solution is given in appendix \ref{Appendix_A}. Interestingly, this is exactly the standard solution with $k\to\tilde k\doteq c_{\rm T}(t) \,k$ therefore, in the presence of a non-trivial propagating speed $c_{\rm T}$, the primordial tensor and scalar spectra at a given scale $k$ are written as \cite{Riotto:2018pcx,Baumann:2014nda,Fumagalli:2016afy}
\begin{equation}
\mathcal P_{\rm T}(k)= \frac{2}{M_p^2\pi^2} \frac{H^2}{c_{\rm T}}\left(\frac{c_{\rm T} k }{aH}\right)^{-2\epsilon_1-\epsilon_1^{\rm T}}
\end{equation}
\begin{equation}
\mathcal P_{\rm S}(k)=\frac{1}{8\pi^2\,M_p^2}\frac{H^2}{\epsilon_1}\left(\frac{k}{aH}\right)^{-2\epsilon_1-\epsilon_2}
\end{equation}

\section{Generalized Consistency Relations}\label{sec.generalcons}
In this section we are going to derive some consistency relations among the inflationary parameters and the propagating speed $c_{\rm T}$.
It is well known that the standard slow roll paradigm of inflation predicts a set of consistency relations that relate the scalar and tensor parameters \cite{Giare:2019snj}. 
As we are going to see, the effects of a non-trivial propagating speed during inflation are encoded in the inflationary parameters and translated into different consistency relations with respect to the standard case. Future detection of the tensor spectrum and a consequent test of these consistency relations can therefore be used to constrain the propagating speed $c_{\rm T}$ testing possible deviations from GR on the inflationary energy scales.

Because of the propagating speed $c_{\rm T}$, the scalar and tensor perturbations now  exit the horizon at different scales. In fact the tensor perturbation will cross the horizon at $c_{\rm T}k=aH$ while the scalar perturbation will cross the horizon\footnote{We are considering the case of a scalar speed $c_{\rm S}=1$} at $k=aH$.
Deriving the primordial spectra, we can compute the scalar and tensor tilts:
\begin{equation}
n_{\rm S}-1\doteq\frac{d\log\mathcal P_{\rm S}}{d\log k}\bigg\rvert_{k=k_*}=-2\epsilon_{1}-\epsilon_2 + O(\epsilon^2) 
\label{ns}
\end{equation}
\begin{equation}
n_{\rm T}\doteq\frac{d\log\mathcal P_{\rm T}}{d\log k}\bigg\rvert_{k=k_*}=-2\epsilon_1-\epsilon^{\rm T}_1 + O(\epsilon^2) 
\label{nt}
\end{equation}
where $k_*$  is the pivot scale and the expressions above hold both for $k_*=aH$ and for $k_*=\frac{aH}{c_{\rm T}}$ at least of corrections of order $\mathcal O(\epsilon^2)$ and therefore negligible. 

As concerns the scalar and tensor amplitudes, also in this case they do not depend drastically on the pivot scale, in fact:
\begin{equation}
\mathcal P_{\rm T}\bigg\rvert_{k_*=\frac{aH}{c_{\rm T}}}= \frac{2}{M_p^2\pi^2} \frac{H^2}{c_{\rm T}}\simeq \frac{2\,H^2}{M_p^2\pi^2} \left(c_{\rm T}\right)^{n_{\rm T}-1}=\mathcal P_{\rm T}\bigg\rvert_{k_*=aH}
\label{Pt}
\end{equation} 
\begin{equation}
    \mathcal P_{\rm S}\bigg\rvert_{k_*=\frac{aH}{c_{\rm T}}}= \frac{\left(c_{\rm T}\right)^{1-n_{\rm S}}}{8\pi^2\,M_p^2}\frac{H^2}{\epsilon_1} \simeq \frac{1}{8\pi^2\,M_p^2}\frac{H^2}{\epsilon_1}=\mathcal P_{\rm S}\bigg\rvert_{k_*=aH}
    \label{Ps}
\end{equation} 
and so the tensor-to-scalar ratio $r\doteq\frac{\mathcal P_{\rm T}(k_*)}{\mathcal P_{\rm S}(k_*)}$ 
\begin{equation}
r\bigg\rvert_{k_*=\frac{aH}{c_{\rm T}}}= 16\epsilon_1\,(c_{\rm T})^{n_{\rm S}-2}\simeq \frac{16\epsilon_1}{c_{\rm T}}\simeq 16\epsilon_1\,\left(c_{\rm T}\right)^{n_{\rm T}-1}=r\bigg\rvert_{k_*=aH}
\label{r}
\end{equation}
In the equations above we have used the fact that we measure $n_{\rm S}\simeq0.96$ \cite{Akrami:2018odb} and we expect $|n_{\rm T}|\ll1$. Note also that we are not interested in a large deviation from the standard GR prescription $c_{\rm T}/c=1$ and that the same results should be obtained computing the scalar and tensor spectra at their respective (different) exit scales. Since we proved that the choice of the pivot scale is not crucial, in this work we will adopt the conventional pivot scale $k_*=aH=0.05\,\rm{Mpc}^{-1}$ unless specified differently.

A first obvious consequence of a non-trivial propagating speed is that the amplitude of the tensor spectrum does not fix anymore the energy scale of inflation directly. In fact in the standard case $\mathcal P_{\rm T}\propto H^2\propto \rho_{\rm{inf}}$ while from Eq.  \eqref{Pt} we see that $\mathcal P_{\rm T} \propto \frac{H^2}{c_{\rm T}}$. 

A more interesting effect of a (slightly) time dependent propagating speed is that the expression for the tensor tilt $n_{\rm T}$ acquires a term $\epsilon^{\rm T}_1$ with respect to the standard case. The sign of $n_{\rm T}$ now depends on the parameter $\epsilon^{\rm T}_{1}$ that quantifies the variation of $c_{\rm T}$ in a Hubble time. In fact, if we consider Eq.\eqref{nt} we see that\footnote{Remember that $\epsilon_1>0$ to ensure the Null Energy Condition.} if during the inflation the propagating speed increases or remains constant in time ($\epsilon^{\rm T}_1\geq0$) the tensor tilt is red ($n_{\rm T}<0$). If instead the propagating speed reduces in time ($\epsilon^{\rm T}_{1}<0$), the sign of $n_{\rm T}$ depends on the magnitude of  $\epsilon^{\rm T}_{1}$. For $-2\epsilon_1<\epsilon^{\rm T}_1<0$  the dismissing is small enough to ensure a negative tensor tilt while for  $\epsilon^{\rm T}_1 < -2\epsilon_1$ the dismissing is translated into a blue tensor tilt $n_{\rm T}>0$. As we will discuss in Sec. \ref{Constraints}, a positive tensor tilt would amplify the PGWs production on small scales and this is why we can use small scale experiments (such as LIGO and VIRGO) to constrain the propagating speed. 

Moreover, as one can see from \eqref{nt} and \eqref{r}, also the usual consistency relation $r=-8n_{\rm T}$ is violated in the presence of a non-trivial propagating speed. In practice, however, there are many ways to violate the consistency relation between $r$ and $n_{\rm T} $ that do not imply a deviation form GR. This means that, if a violation of the consistency relation $r=-8n_{\rm T}$ is observed, we need a way to recognize if such a violation is due to a non-trivial tensor propagating speed during inflation or not. 

As we are going to show we can derive a set of consistency relations among the inflationary parameters and the propagating speed $c_{\rm T}(t)$. For simplicity we suppose that, during inflation, $c_{\rm T}$  increases or decreases linearly with time, so that
\begin{equation}
\ddot c_{\rm T}(t)\simeq 0. 
\label{linear_order}
\end{equation}
In other words, we take into account only the linear term in the Taylor expansion of $c_{\rm T}(t)$. This (reasonable) approximation is not crucial for our results, but simplifies the relations we are going to derive (we discuss scenarios beyond the assumption of linear time evolution for the tensor propagating speed in Appendix \ref{App.B}). 

To relate the propagating speed $c_{\rm T}$ to the inflationary parameters we introduce the scalar and tensor runnings

\begin{equation}
\alpha_{\rm S}\doteq\frac{d\,n_{\rm S}}{d\log k}\bigg\rvert_{k=k_*}=-2\epsilon_1\epsilon_2-\epsilon_2\epsilon_3
\label{as}
\end{equation}
\begin{equation}
\alpha_{\rm T}\doteq\frac{d\,n_{\rm T}}{d\log k}\bigg\rvert_{k=k_*}=-2\epsilon_1\epsilon_2-\epsilon^{\rm T}_1\epsilon^{\rm T}_2
\label{alpha_t}
\end{equation}
because of \eqref{linear_order}, $\epsilon^{\rm T}_2$ can be calculated from its definition \eqref{et2}
\begin{align}
\epsilon^{\rm T}_2\doteq \frac{\dot \epsilon^{\rm T}_1}{H\epsilon^{\rm T}_1}&=\frac{1}{H\epsilon^{\rm T}_1}\frac{d}{dt}\frac{\dot c_{\rm T}}{H c_{\rm T}}=\frac{1}{H\epsilon^{\rm T}_1} \left[\epsilon_1\,\frac{\dot c_{\rm T} }{c_{\rm T}} - \frac{\dot c_{\rm T}^2}{H c_{\rm T}^2}\right] =\epsilon_1-\epsilon^{\rm T}_1
\label{e_2t}
\end{align}
that gives for $\alpha_{\rm T}$
\begin{equation}
\alpha_{\rm T}=-2\epsilon_1\epsilon_2-\epsilon^{\rm T}_1\left(\epsilon_1-\epsilon^{\rm T}_1\right)
\label{alpha_t_1}
\end{equation}
Equations \eqref{ns} \eqref{nt} \eqref{r} \eqref{as} and \eqref{alpha_t_1} can be reversed together to obtain
\begin{align}
&\epsilon_1=\frac{1}{16}\left(r\,c_{\rm T}\right) \label{e1} \\
&\epsilon^{\rm T}_1=-n_{\rm T}-\frac{1}{8}\left(r\,c_{\rm T}\right) \label{eT}\\
&\epsilon_2=1-n_{\rm S}-\frac{1}{8}\left(r\,c_{\rm T}\right) \label{e2}\\
&\epsilon_3=\frac{\alpha_{\rm S}}{n_{\rm S}-1+ 1/8\,\left(r\,c_{\rm T}\right)} -\frac{1}{8}\left(r\,c_{\rm T}\right)  \label{e3}
\end{align}
Using the above equations in $\alpha_{\rm T}$ one obtains
\begin{equation}
\alpha_{\rm T}=n_{\rm T}^2+\frac{5}{128}\left(r\,c_{\rm T}\right)^2 +\frac{1}{8}\left(r\,c_{\rm T}\right)\left[\left(n_{\rm S}-1\right)+\frac{5}{2}n_{\rm T}\right].
\label{cT_alpha_T}
\end{equation}
Equation \eqref{cT_alpha_T} is a consistency relation between $n_{\rm S}$, $n_{\rm T}$, $\alpha_{\rm T}$ and $c_{\rm T}$ that generalizes the usual slow roll relation. Note that we can obtain as many relations as we want; for example, considering also the running of running $\beta_{\rm T}$ 
\begin{equation}\label{Eq.betaT}
\beta_{\rm T}\doteq\frac{d\alpha_{\rm T}}{d\log k}\bigg\rvert_{k=k_*}=-2\epsilon_1\epsilon_2^2-2\epsilon_1\epsilon_2\epsilon_3-\epsilon^{\rm T}_1\left[\left(\epsilon_1-\epsilon^{\rm T}_1\right)^2 +\epsilon_1\epsilon_2-\epsilon^{\rm T}_1\left(\epsilon_1-\epsilon^{\rm T}_1\right)\right]
\end{equation}
it is easy to see that, using \eqref{e1}, \eqref{eT}, \eqref{e2}, \eqref{e3} and solving Eq.\eqref{cT_alpha_T} for $c_{\rm T}$ one obtains a consistency relation $\beta_{\rm T}=\beta_{\rm T}(n_{\rm S}\,,\,\alpha_{\rm S},r\,,\,n_{\rm T}\,,\,\alpha_{\rm T})$. This can be trivial generalized to all orders following the procedure described in \cite{Giare:2019snj} for the standard case. It is, however, more interesting to study some limits of Eq.\eqref{cT_alpha_T} .

\subsection{The limit \boldmath{$\epsilon^{\rm T}_1\to 0$} (i.e. \boldmath{$c_{\rm T}=\rm{Const.}$})}
The limit $\epsilon^{\rm T}_1=0$ describes a constant propagating speed not necessarily equal to the speed of light. Because of \eqref{eT} we have  
\begin{equation}
c_{\rm T}=\frac{-8\,n_{\rm T}}{r}
\label{r_nt_ct=const}
\end{equation}
Using Eq. \eqref{r_nt_ct=const} in the consistency relations \eqref{cT_alpha_T} we obtain
\begin{equation}
\alpha_{\rm T} =n_{\rm T}^2-n_{\rm T}\left(n_{\rm S}-1\right)
\label{RC1}
\end{equation} 
That is the same consistency relation among $n_{\rm T}$, $\alpha_{\rm T}$ and $n_{\rm S}$ than in the standard slow roll case \cite{Giare:2019snj}. Similarly the equation for $\beta_{\rm T}$
\begin{equation}
\beta_{\rm T}=n_{\mathrm{T}}\left(\alpha_{\mathrm{T}}-\alpha_{\mathrm{S}}\right)+\frac{\alpha_{\mathrm{T}}^{2}}{n_{\mathrm{T}}}
\end{equation} 
is the same than the standard slow roll. This mean that if during inflation $c_{\rm T}=\rm{const}\ne1$, the consistency relation between $r$ and $n_{\rm T}$ will be violated but all the other consistency relations will be preserved. If together with $\epsilon^{\rm T}_1=0$ we fix also $c_{\rm T}=1$ (recovering the standard GR prescriptions) the relation $r=-8\,n_{\rm T}$ as well as all the other standard slow roll results will be restored.
\subsection{Limit \boldmath{$c_{\rm T}\to 1$} at the end of inflation}
For completeness we briefly discuss another interesting case in which at the horizon crossing the propagating speed reaches the value $c_{\rm T}\simeq 1$ even with a non vanishing $\epsilon^{\rm T}_1\ne 0$\footnote{This is possible if for example the initial propagating speed was smallest than the speed of light and, at some point, it starts constantly increasing ($\epsilon^{\rm T}_{1}>0$) to reach the value $c_{\rm T}\simeq1$ at the horizon exit.}. In this case we have to simply put $c_{\rm T}=1$ in the Eq. \eqref{cT_alpha_T} obtaining
\begin{equation}
\alpha_{\rm T}=n_{\rm T}^2+\frac{5}{128}\,r^2+\frac{r}{8}\left[\left(n_{\rm S}-1\right)+\frac{5}{2}\,n_{\rm T}\right]
\end{equation}
that is different from the standard slow roll relation \eqref{RC1}. In fact being $\epsilon^{\rm T}_1\ne0$ because of Eq. \eqref{eT} also $n_{\rm T}\ne\frac{-r}{8}$. This means that a time variation of $c_{\rm T}$ can leave a trace even if at the horizon exit the usual GR condition $c_{\rm T}=c=1$ is restored. We conclude that, together with the propagating speed $c_{\rm T}$, another interesting parameter to analyze is $\epsilon^{\rm T}_1$.


\section{Constraints}
\label{Constraints}
So far we have derived a set of consistency relations that generalize the standard slow roll relations introducing a non-trivial tensor propagating speed $c_{\rm T}(t)$. We have shown that the propagating speed can be related to the inflationary parameters which means that they can be used to constrain the propagating speed itself and to test possible deviations from GR at the high energy scales of inflation. 

In this section, we discuss the constraints coming from present cosmological data and imposing the generalized consistency relations that we have derived in Sec. \ref{sec.generalcons}. The theoretical model is calculated using the latest version of the Boltzmann code CAMB \cite{Lewis:1999bs,Howlett:2012mh} and we use the python sampler Cobaya \cite{torrado:2020xyz} to extract cosmological constraints. The posteriors of our parameter space have been explored using the Monte Carlo Markov-Chain (MCMC) sampler developed for CosmoMC \cite{Lewis:2002ah,Lewis:2013hha} and tailored for parameter spaces with a speed hierarchy which also implements the "fast dragging" procedure described in \cite{Neal:2005}. The convergence of the chains obtained with this procedure is tested using the Gelman-Rubin criterium \cite{Gelman:1992zz} and we choose as a threshold for chain convergence $R-1 \lesssim 0.01 $. 
To compare current data with our theoretical model, we employ the Planck's 2018 temperature and polarization likelihood which also includes low multipole data ($\ell < 30$) \cite{Aghanim:2019ame} combined with the lensing likelihood of Planck's 2018 data release based on temperature and polarization lensing reconstruction \cite{Aghanim:2018oex}  and the CMB power spectrum likelihood of Bicep2/Keck Array X (BK15) \cite{Ade:2018gkx}. We report the results coming from our MCMC sampling in Sec. \ref{sec.CMBconstraints}.  

In Sec. \ref{sec.LVconstraints} we will instead focus on the constraints from small scale experiments (namely the LIGO/VIRGO upper limit on the stochastic gravitation waves background, that we denote with LV). 
In fact, for a blue tilted spectrum, the stochastic background of primordial gravitational waves $\Omega_{\rm GW}$ can be strongly amplified on small scales and we can use the small scales experiment data on the stochastic background to constrain the propagating speed and its time variation. Finally in Sec. \ref{sec.combined},  we combine the CMB data and the LIGO/VIRGO bound on the stochastic background to improve the final constraints on the inflationary parameters.

\subsection{Constraints from CMB}\label{sec.CMBconstraints}

\begin{figure}
    \centering
    \includegraphics[width=.9\textwidth]{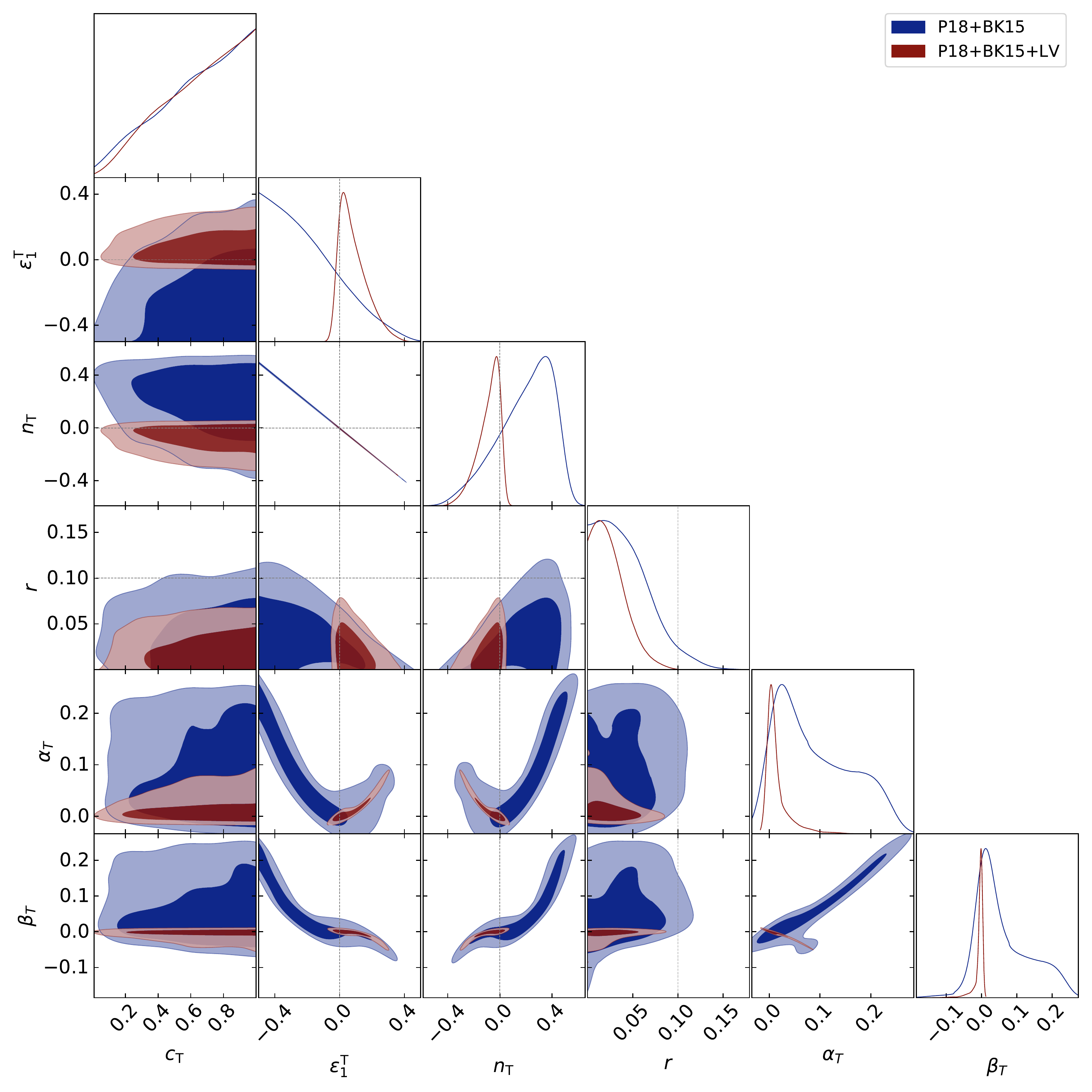}
    \caption{Marginalized 2D and 1D posteriors for the combination of Planck 2018 \cite{Aghanim:2019ame,Aghanim:2018oex} and Biceps/Keck 2015 \cite{Ade:2018gkx} data for the parameters of the tensor spectrum and their combination with the LIGO/VIRGO upper limit on the stochastic background amplitude \cite{LIGO_SGWB-2017,LIGO_SGWB-2019} (P18+BK15+LV).}
    \label{fig:figure2}
\end{figure}

\begin{figure}
    \centering
    \includegraphics[width=0.9\textwidth]{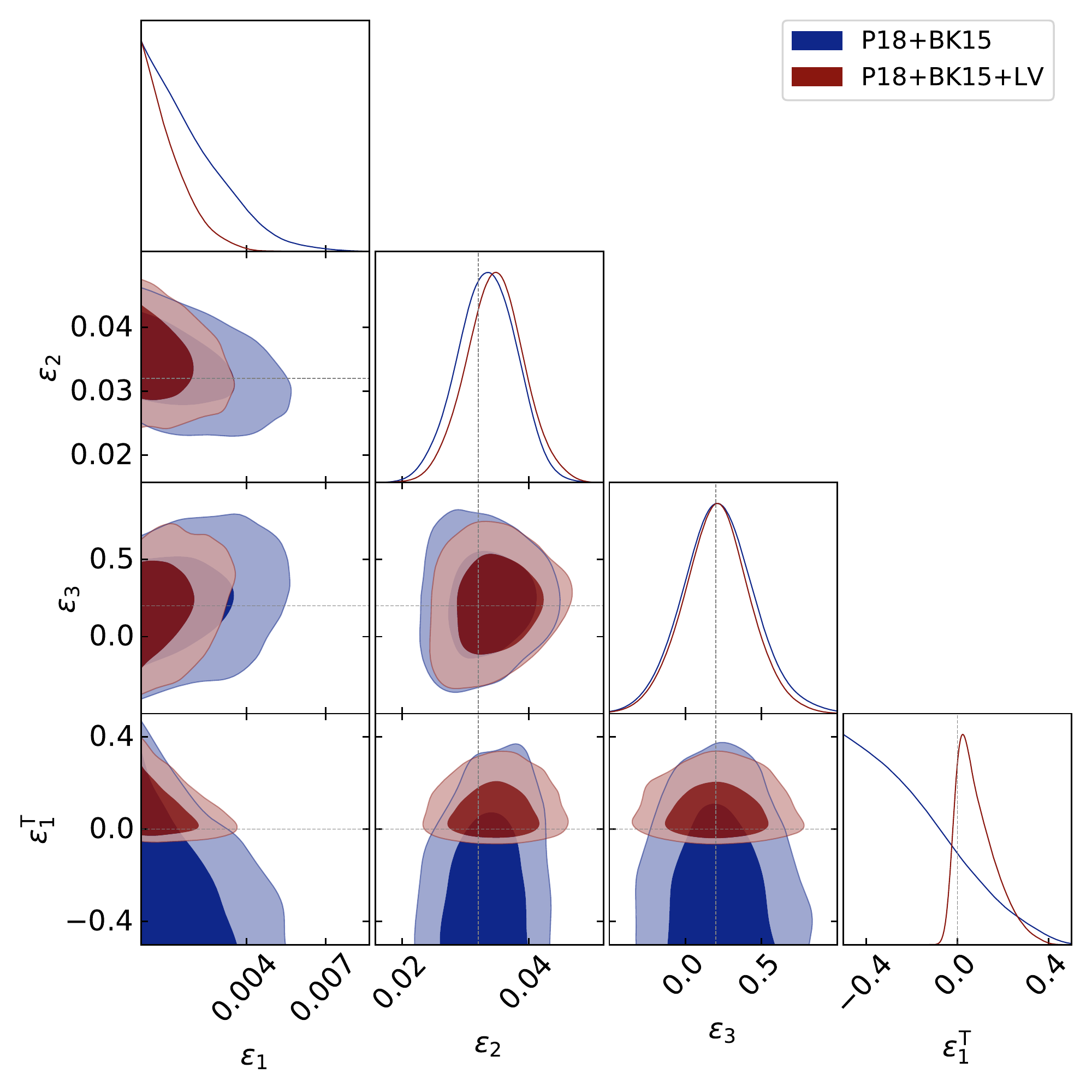}
    \caption{Marginalized 2D and 1D posterior for the combination of Planck 2018 \cite{Aghanim:2019ame,Aghanim:2018oex} and Biceps/Keck 2015 \cite{Ade:2018gkx} data (P18+BK15) for the first and second order slow parameters and their combination with the LIGO/VIRGO upper limit on the stochastic background amplitude \cite{LIGO_SGWB-2017,LIGO_SGWB-2019} (P18+BK15+LV). }
    \label{fig:figure3}
\end{figure}

\begin{table}[h]
\begin{center}
\renewcommand{\arraystretch}{1.4}
\begin{tabular}{c@{\hspace{1 cm}}@{\hspace{1 cm}} c}
\hline
\textbf{Parameter}                    & \textbf{Prior}\\
\hline\hline
$\Omega_{\rm b} h^2$         & $[0.005\,,\,0.1]$\\
$\Omega_{\rm c} h^2$       & $[0.001\,,\,0.99]$\\
$100\,\theta_{\rm {MC}}$             & $[0.5\,,\,10]$\\
$\tau$                       & $[0.01\,,\,0.8]$\\
$\log(10^{10}A_{\rm S})$         & $[1.61\,,\,3.91]$\\
$n_S$                        & $[0.8\,,\, 1.2]$\\
$c_{\rm T}$ &$[0.01\,,\,1]$\\
$16\,\epsilon_1$ & $[0 \,,\,1]$\\ 
$\epsilon_1^{\rm T}$ & $[-0.5\,,\,0.5]$\\
$\epsilon_3$ & $[-0.5\,,\,1]$\\
\hline\hline
\end{tabular}
\end{center}
\caption{List of the parameters used in the MCMC sampling and their external flat priors assumed in Sec. \ref{sec.CMBconstraints}. In Sec. \ref{sec.combined} we sampled the same parameters with the same external priors except for $\epsilon^{\rm T}_1$ on which we also impose a Half-Gaussian prior to include LIGO/VIRGO data on the stochastic background \cite{LIGO_SGWB-2017,LIGO_SGWB-2019}}
\label{priors}
\end{table}

\begin{table*}[!hbtp]
	\centering
	\begin{tabular}{lcccc}
		\toprule
		\horsp    \vertsp P18+BK15 \vertsp  P18+BK15+LV  \\
		\hline\hline
		\morehorsp
		$\Omega_{\mathrm{b}} h^2$     \vertsp $ 0.02242\pm 0.00015 $   \vertsp  $ 0.02241\pm 0.00015 $       \\  
		$\Omega_{\mathrm{c}} h^2$     \vertsp $0.1200\pm 0.0012 $  \vertsp  $0.1200\pm 0.0012 $        \\  
		$ \tau $  	\vertsp $0.0566\pm 0.0076 $	    \vertsp  $0.0564\pm 0.0079 $                 \\	
		$\ln(10^{10} A_\mathrm{S}) $  \vertsp $ 3.051\pm 0.015$   \vertsp  $ 3.050\pm 0.016$     \\  
     	$ r $  \vertsp $ < 0.0961 $	\vertsp  $ < 0.0599 $		             \\  
		$n_{\rm S} $  	\vertsp $ 0.9645\pm 0.0044$	     \vertsp  $ 0.9646\pm 0.0044 $             \\
		$\alpha_{\rm S} $  \vertsp $-0.0067\pm 0.0067 $     \vertsp  $  -0.0069\pm 0.0069 $        \\
		$n_T$    \vertsp $ 0.20^{+0.27}_{-0.13}$        \vertsp  $ -0.084^{+0.10}_{-0.047} $          \\
     	$\alpha_{\rm T} $  \vertsp $0.087^{+0.049}_{-0.098}$     \vertsp  $ 0.0141^{+0.0035}_{-0.021}$        \\
		$c_{\rm T}$   \vertsp $ >0.178   $     \vertsp  $ > 0.219 $                \\       
		$\epsilon^{\rm T}_1$   \vertsp $ < 0.203 $     \vertsp  $ 0.082^{+0.047}_{-0.11} $                \\       
		$ \chi^2 $   \vertsp $ 3530 $     \vertsp  $3530 $                \\
		\bottomrule
	\end{tabular}
    \caption{Constraints on parameters are at 1$\sigma$ level ($68\%$ C.L.) while upper bounds are at 2$\sigma$ ($95\%$ C.L.) for the full Planck 2018 likelihood \cite{Aghanim:2019ame,Aghanim:2018oex} and Biceps/Keck 2015 B-mode \cite{Ade:2018gkx} likelihood with and without the inclusion of the prior on $\epsilon_1^{\rm T}$ coming from LIGO/VIRGO data \cite{TheLIGOScientific:2016dpb}}
    \label{Tab.CMBresults}
\end{table*}

In this subsection we present the results of our MCMC analysis. Let us start by noting that the Boltmann integrator CAMB \cite{Lewis:1999bs,Howlett:2012mh} employs the standard power law parametrization of the primordial scalar and tensor power spectra \emph{i.e.} :
\begin{align}
    \mathcal{P}^{\rm CAMB}_{\rm S}(k) & = A_{\rm S}^{\rm CAMB}\left(\frac{k}{k_{\star, \rm S}} \right)^{n_{\rm S}-1 + \frac{1}{2}\,\alpha_{\rm S}\,\log(k/k_{\star,\rm S})} \\  
    \mathcal{P}^{\rm CAMB}_{\rm T}(k) & = A_{\rm T}^{\rm CAMB}\left(\frac{k}{k_{\star, \rm T}} \right)^{n_{\rm T} + \frac{1}{2}\,\alpha_{\rm T}\,\log(k/k_{\star,\rm T})}
\end{align}
where $k_{\star, \rm T}$ and $k_{\star, \rm S}$ are the tensor and scalar pivot scale and the tensor-to-scalar ratio is defined as $r^{\rm CAMB} = \mathcal{P}^{\rm CAMB}_{\rm T}(k_{\star, \rm T})/\mathcal{P}^{\rm CAMB}_{\rm S}(k_{\star, \rm S}) $. 
While the inclusion of a non-trivial tensor propagating speed leaves unchanged the scalar spectrum, it impacts the tensor spectrum by rescaling its amplitude of a factor $ c_T^{n_{\rm T} - 1} $. We therefore modify CAMB in order to include this correction by rescaling $r^{\rm CAMB}$ accordingly (\emph{i.e.} $r = r_{0.05} = r^{\rm CAMB} c_{\rm T}^{n_{\rm T} -1}$) and calculating the amplitude of the spectra at the same pivot scale $k_{\star, \rm T} = k_{\star, \rm S} = aH = 0.05\ \rm Mpc^{-1} $. This choice ensures that $r$ is calculated to a well-defined scale and allows our constraints to be directly compared with the results reported by the Planck Collaboration \cite{Akrami:2018odb,Aghanim:2018eyx}. In our MCMC analysis we consider the six parameters of the standard $\Lambda$CDM model i.e. the baryon $\omega_{\rm b}\doteq\Omega_{\rm b}\,h^2$  and cold dark matter  $\omega_{\rm c}\doteq\Omega_{\rm c}\,h^2$ energy densities, the angular size of the horizon at the last scattering surface $\theta_{\rm MC}$, the optical depth $\tau$, the amplitude of primordial scalar perturbation $\log(10^{10}\,A_{\rm S})$ and the scalar spectral index $n_{\rm S}$. As discussed in the introduction, the inclusion of the tensor and scalar runnings may significantly enhance the constraints on the parameters describing tensor spectra from current data. Therefore, along with the six standard $\Lambda\rm CDM$ parameters, we also include in our analysis the scalar running $\alpha_{\rm S}$, the tensor-to-scalar ratio $r$, the tensor spectral index $n_{\rm T}$, the tensor running $\alpha_{\rm T}$, the propagating speed $c_{\rm T}$ and the parameter $\epsilon_1^{\rm T}$ that quantifies its time variation per Hubble time.  Instead of directly sampling these parameters (as it is commonly done, see e.g \cite{Akrami:2018odb,Aghanim:2018eyx}) we choose to do the MCMC sampling using, along with the standard $\Lambda$CDM parameters, the following four $\{c_T, 16\epsilon_1,\epsilon_1^T,\epsilon_3\}$ and to derive the value of the tensor and scalar runnings from the generalized consistency relations introduced in Sec. \ref{sec.generalcons}. The flat priors\footnote{Note that in our MCMC sampling we are considering only the parameter space of subluminal velocities. We discuss superluminal velocities in appendix \ref{app.C}.} on our parameter space are reported in tab \ref{priors}.

In Table \ref{Tab.CMBresults} we show the constraints on the parameters from the combination of Planck and Biceps/Keck data while in Fig.\ref{fig:figure2} we report their $68\%$ and $95\%$ contour plots. 
A first aspect we would like to stress is that our results confirm that a non-trivial time-dependent propagating speed does not alter the constraints on the scalar parameters from the Planck data (which assumes $c_T = 1$) as expected from our theoretical discussion.

As concerns the inflationary tensor parameters, the tensor propagating speed $c_{\rm T}$ is only weakly constrained with the 95\% C.L. contours showing a preference for $c_{\rm T}\gtrsim 0.18$. This is expected since the CMB data only constrain the amplitude of tensor perturbations $A_T = r\,A_{\rm S} =16\epsilon_1\,A_{\rm S}\,c_{\rm T}^{n_{\rm T} - 1}$. Then Planck data are only able to bound the product  $\epsilon_1/ c_{\rm T}$ and since they prefer a tensor amplitude consistent with zero this leads to a weakly constrained propagating speed of tensor perturbations; only an upper bound can be placed on the tensor-to-scalar ratio $r<0.096$ at 95\% C.L. Nevertheless we can derive the upper bound $\epsilon_1^{\rm T}<0.203$ at $95\%$ C.L. on the parameter that quantifies the time dependence of $c_{\rm T}$. The fact that the region $\epsilon_1^{\rm T}<0$ is essentially unconstrained from the Planck data is translated into the fact that the tensor tilt can assume large positive values as well as the tensor running $\alpha_{\rm T}$. 

We note that the bound we derive on the tensor-to-scalar ratio is $\sim 60\%$ worse with respect to the results obtained from a combination of Planck and Biceps data without considering the runnings of the tensor spectrum. Conversely, the bound on the tensor spectral index $n_T$ is significantly improved. In particular, $ -0.23 \leq n_T \leq 0.54 $ at $95\%$ C.L. showing an improvement of a factor of 2 in the negative tail and a factor of 5 improvement in the positive tail in place of the Planck results of $-0.55\leq n_T \leq 2.54$. This situation is again a direct consequence of considering a non-vanishing tensor running and imposing the generalized consistency relation \eqref{cT_alpha_T}. 
When $\alpha_T$ is non-zero the tensor spectrum acquires a term $\sim \alpha_{\rm T}\,\log^2 k$ leading to a growth on small scales (high $k$). The freedom in $n_{\rm T}$ is so partially transferred to $\alpha_{\rm T}$ that it results to be almost the same order of magnitude as $n_{\rm T}$. Moreover, from Eq. \eqref{Eq.betaT} one can also derive a constraint on the second-order tensor running $\beta_{\rm T}$ that we found to be $\beta_{\rm T}=0.060^{+0.046}_{-0.093}$ at 68\% C.L. (i.e. again of almost the same order than $n_{\rm T}$ and $\alpha_ {\rm T}$)\footnote{These results are consistent with the relation $\beta_{\rm T}\simeq 2\,n_{\rm T}^3 \simeq 2 \alpha_{\rm T}^{3/2}$, discussed in Sec. \ref{sec.LVconstraints}.}
This shows that also the constraints from the CMB data can be sensitive to the higher-order terms in the primordial spectra, enforcing the importance of a proper parametrization to correctly describe their small scale behavior.
Indeed, such large positive values of $n_T$ (and its runnings) would amplify the production of PGWs at small scales and possibly lead to an amplitude $\Omega_{\rm GW}$ higher than the current LIGO/VIRGO bound at $k{\rm LV}$. As we describe in detail in the following section, the LIGO/VIRGO constraints on the stochastic background can be used to put tight constraints on the small scale behavior of the tensor spectrum.

For completeness we also report the bound on the standard slow roll parameters that can be derived accordingly to the consistency relation derived in Sec. \ref{sec.generalcons}. We obtain the following constraints from the combination P18+BK15:
\begin{align}
    &\epsilon_1 < 0.0046   &(95\%\ \rm{C.L}) \\
    &\epsilon_2 = 0.0334\pm 0.0046 &(68\%\ \rm{C.L}) \\
    &\epsilon_3 = 0.22\pm 0.23  &(68\%\ \rm{C.L}) 
\end{align}
in very good agreement with the results of the Planck Collaboration \cite{Aghanim:2018eyx}.
We show the 2D marginalized contour plots and 1D marginalized posterior distributions of these parameters in Fig.\ref{fig:figure3}.

\subsection{Constraints from small scale experiments on Gravitational Waves}\label{sec.LVconstraints}
In this subsection, we are going to derive constraints on the inflationary parameters discussed in this work from small scales data.
If during inflation the propagating speed of gravitational waves decreases enough (i.e. if $\epsilon^{\rm T}_1$ is negative enough), the tensor tilt can become blue amplifying the Primordial Gravitational Waves production on small scales. As we stated in the introduction, small scale experiments on gravitational waves such as LIGO/VIRGO and, in the future, LISA and Einstein Telescope (ET), are sensitive to the stochastic background, $\Omega_{\rm{GW}}$ and can be used to improve the constraints on the inflationary parameters. In particular Eq. \eqref{blue_tilt} provides a rough estimation of the upper bounds we can set on the blue tensor tilt from small scale experiments, see Fig. \ref{fig:figure1}. However Eq. \eqref{blue_tilt} has been derived assuming that the power law expansion holds from the CMB scales all the way up to the small scales probed by the ground based interferometers that are separated from the CMB scales by a factor of $10^{18}$ in $k$. We have already said that on such small scales the higher order corrections due to the tensor runnings can be non negligible and that should be included in the analysis \cite{Giare:2020vhn}. Therefore in this subsection we generalize the parametrization of the primordial tensor spectrum to the following expansion \cite{Zarei:2014bta,Giare:2019snj}:
\begin{equation}
\mathcal P_{\rm T}(k) =  r\,A_{\rm S}\,\left(\frac{k}{k_*}\right)^{n_{\rm T}(k_*) + \sum_{n=1}^{\infty} \frac{\alpha_n^{\rm T}(k_*)} {(n+1)!} \left[\log\left(\frac{k}{k_*}\right)\right]^n}
\label{Tensor}
\end{equation} 
We recall that the amplitude of the primordial scalar spectrum is measured to be $A_{\rm s}\equiv \mathcal P_{\rm s}(k_*)\simeq 2\times 10^{-9}$ \cite{Akrami:2018odb}. Here we adopt the notation\footnote{In this notation $\alpha_{\rm T}\equiv \alpha^{\rm T}_{1}$ and $\beta_{\rm T}\equiv \alpha^{\rm T}_{2}$}:
\begin{equation}
\alpha_n^{\rm T}(k_*)\doteq \left( \frac{d}{d\log k}\right)^n\,n_{\rm T}(k)\bigg \rvert _{k=k_*}
\end{equation}
for the $n$-order tensor running\footnote{In what follows we will avoid to specify that the spectral tilt and the runnings are computed on the pivot scale $k_*$ and, simplifying  the notation, we will only write $n_{\rm T}$ and $\alpha_n^{\rm T}$.}. 

\begin{figure}[h]
    \centering
    \includegraphics[width=1\textwidth]{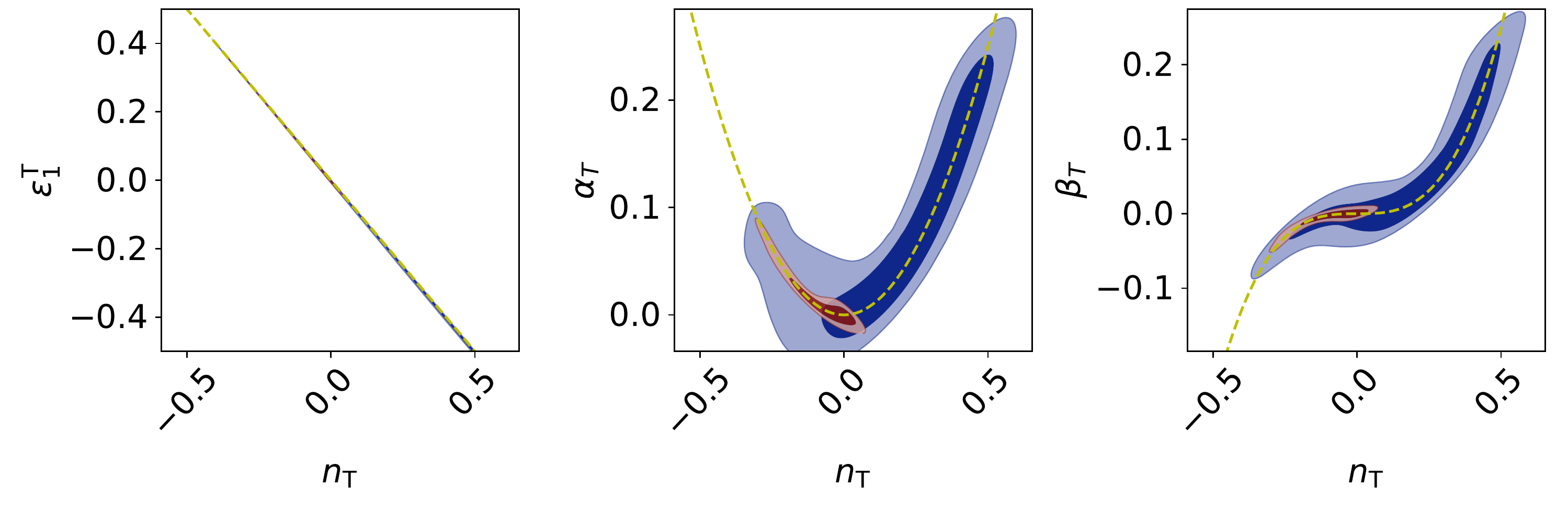}
    \caption{Marginalized 2D posterior in the planes ($n_{\rm T}\,,\,r$) and ($\alpha_{\rm T}\,,\,r$). The blue contours are derived from the combination of Planck 2018 \cite{Aghanim:2019ame,Aghanim:2018oex} and Biceps/Keck 2015 \cite{Ade:2018gkx} data (see Sec. \ref{sec.CMBconstraints}) while the red contours include also the LIGO/VIRGO data on the stochastic background \cite{LIGO_SGWB-2017,LIGO_SGWB-2019} (see Sec. \ref{sec.combined}). The yellow dashed lines represent the relations \eqref{estimation1} and \eqref{estimation2} we used to derive the small scale constraints in Sec. \ref{sec.LVconstraints}.}
    \label{fig:figure4}
\end{figure}

In order to estimate the higher order contributions given by the sum \eqref{Tensor}, we  work under the following assumption: we consider the tensor parameters dominated by the time variation of the propagation speed through the parameter $\epsilon^{\rm T}_1$ in such a way that:
\begin{equation}
n_{\rm_{T}}=-2\epsilon_1-\epsilon^{\rm T}_1\simeq -\epsilon^{\rm T}_1
\label{estimation1}
\end{equation}
and consequently because of Eq. \eqref{e_2t}
\begin{equation}
\alpha_n^{\rm T}\doteq\left(\frac{d}{d\log k}\right)^n n_{\rm_{T}}\simeq n!\,\left(-\epsilon^{\rm T}_1\right)^{n+1} \simeq n!\, \left(n_{\rm T}\right)^{n+1}
\label{estimation2}
\end{equation}
This approximation is in great accordance with the results derived in the previous section as it is possible to see from Fig. \ref{fig:figure4}. In the left panel we plot the constraints in the plane ($n_{\rm T}$ , $\epsilon_1^{\rm T}$) while in the middle and right panels of the same figure we plot the constraints on the first two runnings (i.e. $\alpha^{\rm T}_1\equiv\alpha_{\rm T}$ and $\alpha_2^{\rm T}\equiv\beta_{\rm T}$) in the planes ($n_{\rm T}$ , $\alpha_{\rm T}$) and  ($n_{\rm T}$ , $\beta_{\rm T}$), respectively. As one can see from the left panel the linear relation \eqref{estimation1} between $n_{\rm T}$ and $\epsilon_1^{\rm T}$ is confirmed and the impact of the parameter $\epsilon_1$ is in fact negligible. The middle and right panels, instead validate the relation \eqref{estimation2} between the runnings and the tensor tilt (or equivalently between the runnings and $\epsilon^{\rm T}_1$). In fact we see that $\alpha_{\rm T}\simeq \left(n_{\rm T}\right)^2\simeq \left(\epsilon_1^{\rm T}\right)^2$ while $\beta_{\rm T}\simeq 2\, \left( n_{\rm T} \right)^3\simeq 2\,\left(-\epsilon_1^{\rm T}\right)^3$, which is exactly what we expect from Eq. \eqref{estimation2}. Therefore when $\epsilon^{\rm T}_1$ is negative, not only is the tensor tilt blue but also the runnings are positive. This amplifies the PGWs production on small scales allowing us to further improve the constraints on the inflationary parameters. 

At the end of this section we will come back to further discuss the validity of our approximation. 

Since we are going to constrain the region of the parameter space $\epsilon_1^{\rm T}<0$ it is convenient to use $-\epsilon_1^{\rm T}=|\epsilon_1^{\rm T}|$. Putting \eqref{estimation1} and \eqref{estimation2} into \eqref{Tensor}, we can estimate the sum 
\begin{equation}
\Omega_{\mathrm{GW}}(k) = \frac{r\,A_{\rm S}} {24 z_{\mathrm{eq}}} \left(\frac{k}{k_*}\right)^{-\frac{\log\left(1-|\epsilon^{\rm T}_1|\log\left(\frac{k}{k_*}\right)\right)}{\log\left(\frac{k}{k_*}\right)}}.
\label{Omega}
\end{equation}
As one can see from Eq. \eqref{Omega}, on the generic ultra-high $\tilde k\gg k_*$ the spectrum is well defined if $|\epsilon_1^{\rm T}|\lesssim 1 /\log(\tilde k/k_*)$. More precisely: if $|\epsilon_1^{\rm T}|\ll 1/ \log(\tilde k/k_*)$ the spectrum is essentially flat $\Omega_{\rm GW}\simeq r\,A_{\rm s}/24 z_{\rm eq}$ while if $|\epsilon_1^{\rm T}|\simeq 1/ \log(\tilde k/k_*)$ the spectrum is still flat for $k<\tilde k$, but it exponentially grows at $k\sim \tilde k$.

\begin{figure}[h]
    \centering
    \includegraphics[width=0.5\textwidth]{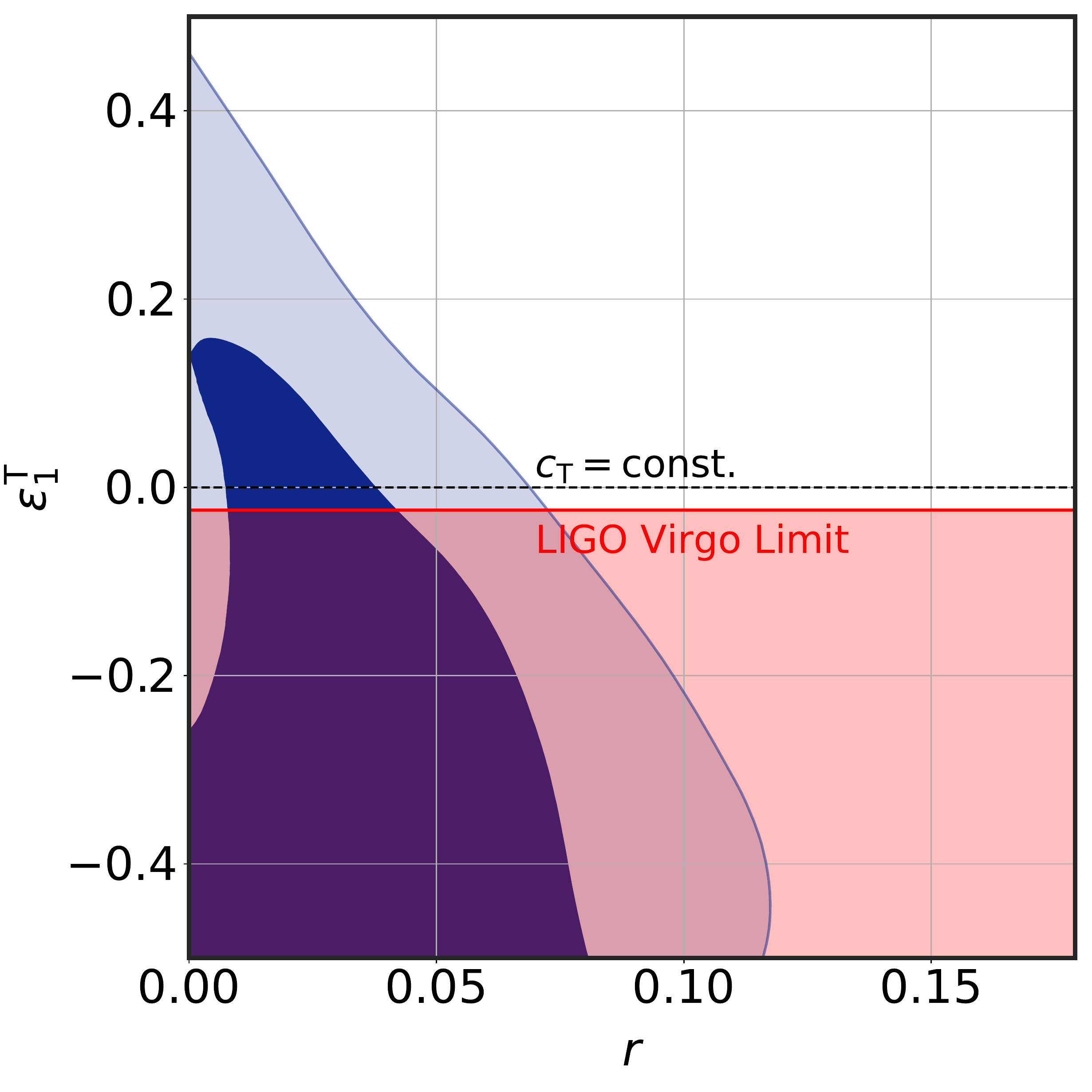}
    \caption{Marginalized 2D posterior for the combination of Planck 2018 \cite{Aghanim:2019ame,Aghanim:2018oex} and Biceps/Keck 2015 \cite{Ade:2018gkx} data in the plane ($r,\,\,\epsilon^{\rm T}_1$). The red region is excluded by the LIGO/VIRGO data on the stochastic background of GWs (see Sec. \ref{sec.LVconstraints}).}
    \label{fig:figure5}
\end{figure}

Here we derive a cutoff on $\epsilon_1^{\rm T}$ simply demanding the spectrum to be well defined \emph{at least} from the CMB scales all the way up to the ultra-high $k$ probed by gravitational detectors and matching the LIGO/VIRGO constraints. In fact, we recall that in the frequency range $f\in\left(20\,\rm{-}\,85.8\right)$ Hz, which corresponds to the wave-number range  $k_{\rm LV} \in \left(1.3\,\rm{-}\,5.5\right) \times 10^{16} \,\rm{Mpc}^{-1}$, the LIGO and VIRGO data set an upper bound on the stochastic background given by Eq.\eqref{LVlimit}. Interestingly, reversing Eq. \eqref{Omega}
\begin{equation}
|\epsilon^{\rm T}_1|=\frac{1-\frac{r\,A_{\rm S}}{24z_{\rm eq} \Omega_{\mathrm{GW}}(k)\,}}{\log\left(\frac{k}{k_*}\right)}.
\label{et_Omega}
\end{equation}  
the LIGO/VIRGO limit on the stochastic background can be translated into a lower bound on $\epsilon_1^{\rm T}$
\begin{equation}
\epsilon^{\rm T}_1\geq - \frac{1-\frac{r\,A_{\rm S}}{24z_{\rm eq}\,\Omega_{\mathrm{GW}}(k_{\rm LV})}}{\log\left(\frac{k_{LV}}{k_*}\right)} \simeq - 0.0249 + \left(3.5\times 10^{-9}\right)r
\label{et_limit}
\end{equation}
that is almost insensitive to the value of the tensor-to-scalar ratio $r$. Equivalently Eq. \eqref{et_limit} puts a stringent upper limit on the blue tensor tilt
\begin{equation}
n_{\rm T}\lesssim 0.025 
\label{nt_LV_WR}
\end{equation}
We plotted the LIGO/VIRGO limit on $\epsilon_1^{\rm T}$ in Fig. \ref{fig:figure5}. As one can see comparing the upper bound \eqref{nt_LV_WR} with that plotted in Fig. \ref{fig:figure1}, once that higher order corrections (i.e. the tensor runnings) are included in the analysis we can improve the final constraints of more than 1 order of magnitude.
Note also that the constraints on $\epsilon^{\rm T}_1$ can be translated into constraints on $c_{\rm T}$ since $\epsilon_1^{\rm T}$ quantifies how the propagating speed changes with respect to the scale. To see this, since here we are focusing only on the linear terms assuming that $\ddot c_{\rm T}\simeq 0$, we can consider a simple toy model where the propagating speed constantly decreases for almost all the e-fold of inflation in such a way that the equation of motion reads
\begin{equation}
c_{\rm T}(t)-c_{\rm T}(t_i)\doteq\int_{t_i}^{t}\dot c_{\rm T}\,dt =\dot c_{\rm T} \left(t-t_i\right)=\epsilon^{\rm T}_1 c_{\rm T}(t)\,\Delta N
\end{equation} 
where $\Delta N=H\Delta t$ is the total number of e-fold between the initial time $t_i$ (when $c_{\rm T}$ starts to decrease) and the time $t$. In this case $c_{\rm T}$ is given by 
\begin{equation}
c_{\rm T}=\frac{c_{\rm T}(t_i)}{1-\epsilon^{\rm T}_1 \Delta N}.
\end{equation}
Assuming $c_{\rm T}(t_i)=1$ and $\Delta N\simeq 60$, the LIGO/VIRGO constraint on $\epsilon^{\rm T}_1$ implies\footnote{We want to stress that this example is used to show that constraints on $\epsilon_1^{\rm T}$ can be translated into constraints on $c_{\rm T}$ assuming that we know how the tensor speed evolves during inflation.
However to derive our final results (shown in Table \ref{Tab.CMBresults}) we did not assume any specific evolution. Note also that in appendix \ref{app.D} we discuss the consistency between our final results and the current small scale measurement of $c_{\rm T}$.} 
\begin{equation}
c_{\rm T} \gtrsim 0.4
\end{equation}
that is consistent with the 2D marginalized posteriors shown in Fig.\ref{fig:figure2} where values of $c_{\rm T}$ smaller than 0.4 times the speed of light seem to be disfavored, at least within the 68\% C.L. contours.

As concerns the next generation of gravitational waves experiments, LISA and ET are expected to have a sensitivity to the stochastic background $\Omega_{\rm{GW}}(k_{\rm Lisa})\simeq 1\times10^{-12}$ on scales $k_{\rm Lisa}\approx 1\times10^{13}\,\rm{Mpc}^{-1}$ \cite{Bartolo:2016ami} and $\Omega_{\rm{GW}}(k_{\rm ET})\simeq 3\times10^{-13}$ on scales $k_{\rm ET}\approx 5\times 10^{15}\,\rm{Mpc}^{-1}$ \cite{Maggiore:2019uih}, respectively. Considering the higher-order corrections in $\mathcal P_{\rm T}(k)$, we see that the improvement in sensitivity expected from LISA and ET is not translated into constraining power on $\epsilon_1^{\rm T}$ and consequently on the tensor tilt at the CMB scales \footnote{In this model, the constraints on $n_{\rm T}(k_*)$ expected by future experiments are $n_{\rm T}(k_*) \lesssim 0.032$ for LISA and $n_{\rm T}(k_*)\lesssim 0.025$ for ET.}. This result seems to contradict the common intuition but the key aspect here is scale-dependence. Assuming the generalized tensor spectrum of Eq. \eqref{Tensor}, we can define a \textit{scale-dependent} tensor tilt $n_{\rm T}(k)$

\begin{equation}
n_{\rm T}(k)\doteq n_{\rm T}(k_*) + \underbrace{\sum_{n=1}^{\infty} \frac{\alpha^{\rm T}_{n}}{(n+1)!} \left[\log(k/k_*)\right]^n}_{\doteq S(k)}
\label{nT(k)}
\end{equation}
in such a way that we can always derive constraints by $\Omega_{\rm GW}$, trivially generalizing Eq. \eqref{blue_tilt} for the scale-dependent case as
\begin{equation}
n_{\rm T}(k) <\frac{\ln\left(\frac{24\,z_{\rm eq}\,\Omega_{\rm GW}(k)}{r\,\mathcal P_{\rm S}(k_*)}\right)}{\ln\left(\frac{k}{k_*}\right)}
\label{Blue_tilt(k)}
\end{equation}
with $n_{\rm T}(k)$ given by \eqref{nT(k)}. Note that the improvement in the sensitivity expected by LISA and ET is \textit{again} translated into an improvement in the constraint on $n_{ \rm T}(k)$ (the same improvement shown in fig.\ref{fig:figure1}), but now these constraints must be referred to the tensor tilt evaluated at different scales $k$: $n_{\rm T}(k_{\rm ET})\lesssim n_{\rm T}(k_{\rm LISA})< n_{\rm T}(k_{\rm LV})$.

Therefore the improvement in the constraints expected from LISA and ET is not trivially translated into an improvement in the constraints on the tensor tilt on the CMB scales. In fact the constraints on a given scale $k$  are related to the constraints on the CMB scales $k_*$  through  the sum $S(k)$ that carries information about the scale dependence of the specific model \footnote{The scale-dependence is encoded in the runnings $\{\alpha_n^{\rm T}\}$ that define the shape of $n_{\rm T}(k)$ relating its value on the CMB scales with its value on the generic scale $k$ by Eq. \eqref{nT(k)}.}. In the inflationary model considered here, the constraints on $n_{\rm T}(k_*)$ remain almost the same for the three experiments. Indeed while  $n_{\rm T}(k_{\rm ET})\lesssim n_{\rm T}(k_{\rm LISA})< n_{\rm T}(k_{\rm LV})$ it is also true that $S(k_{\rm LV})> S(k_{\rm ET})> S(k_{\rm LISA})$ and the two terms in Eq. \eqref{nT(k)} compensate each other leaving almost the same freedom on the CMB scales for $n_{\rm T}(k_*)$.

Before concluding this section, we want to briefly come back on the approximations \eqref{estimation1} and \eqref{estimation2} on which our results are based. Even if we have already shown that the analysis performed in the previous section confirms their validity, it is worthwhile to additionally prove their robustness. The shape of the tensor tilt plotted in Fig. \ref{fig:figure4} and, in general, the validity of our approximation can be further understood as follows: using Eq. \eqref{r}, we see that the value of $\epsilon_1$ is fixed by the value of $c_{\rm T}$ and $r$:
\begin{equation}
\epsilon_1=\frac{r}{16}c_{\rm T}\lesssim \frac{r}{16}
\end{equation}     
where in the last inequality we have used that $c_{\rm T}\lesssim 1$. From the CMB data we know that $r$ is constrained to be very small, if for example we fix the tensor-to-scalar ratio to $r\sim 10^{-2}$, we immediately see that $\epsilon_1\sim 10^{-4}$ and\footnote{Using Eq. \eqref{ns} and the fact that  $n_{\rm s}\simeq 0.96$ \cite{Akrami:2018odb}} $\epsilon_2\sim 10^{-2}$. So for $|\epsilon^{\rm T}_1|\sim 10^{-2}$ (i.e. the order of the limit we derived from the LIGO and VIRGO data), comparing the terms involved in the generic $n$ order running,
\begin{align}
&|\epsilon^{\rm T}_1|\left(\epsilon_1\right)^n\sim \epsilon_2 \left(\epsilon_1\right)^n\sim 10^{-2\left(2n+1\right)} \\
&\epsilon_1\left(\epsilon_2\right)^n\sim\epsilon_1\left(|\epsilon^{\rm T}_1|\right)^n\sim 10^{-2\left(n+2\right)}\\
&\left(|\epsilon^{\rm T}_1|\right)^{n+1}\sim 10^{-2\left(n+1\right)}
\end{align} 
we find that $\alpha^{\rm T}_{n}\simeq n! \left(-\epsilon^{\rm T}_1\right)^{n+1}$ unless corrections at least 2 order of magnitude smaller. The approximation is even better for smaller $r$ while it is trivial to see that it is still valid for the whole range of $r$ explored in our MCMC analysis as Fig. \ref{fig:figure4} confirms. This proves the robustness of our results, definitively.

\subsection{Combined constraints from CMB and Small scale experiments}\label{sec.combined}
\begin{figure}[h]
    \centering
    \includegraphics[width=0.9\textwidth]{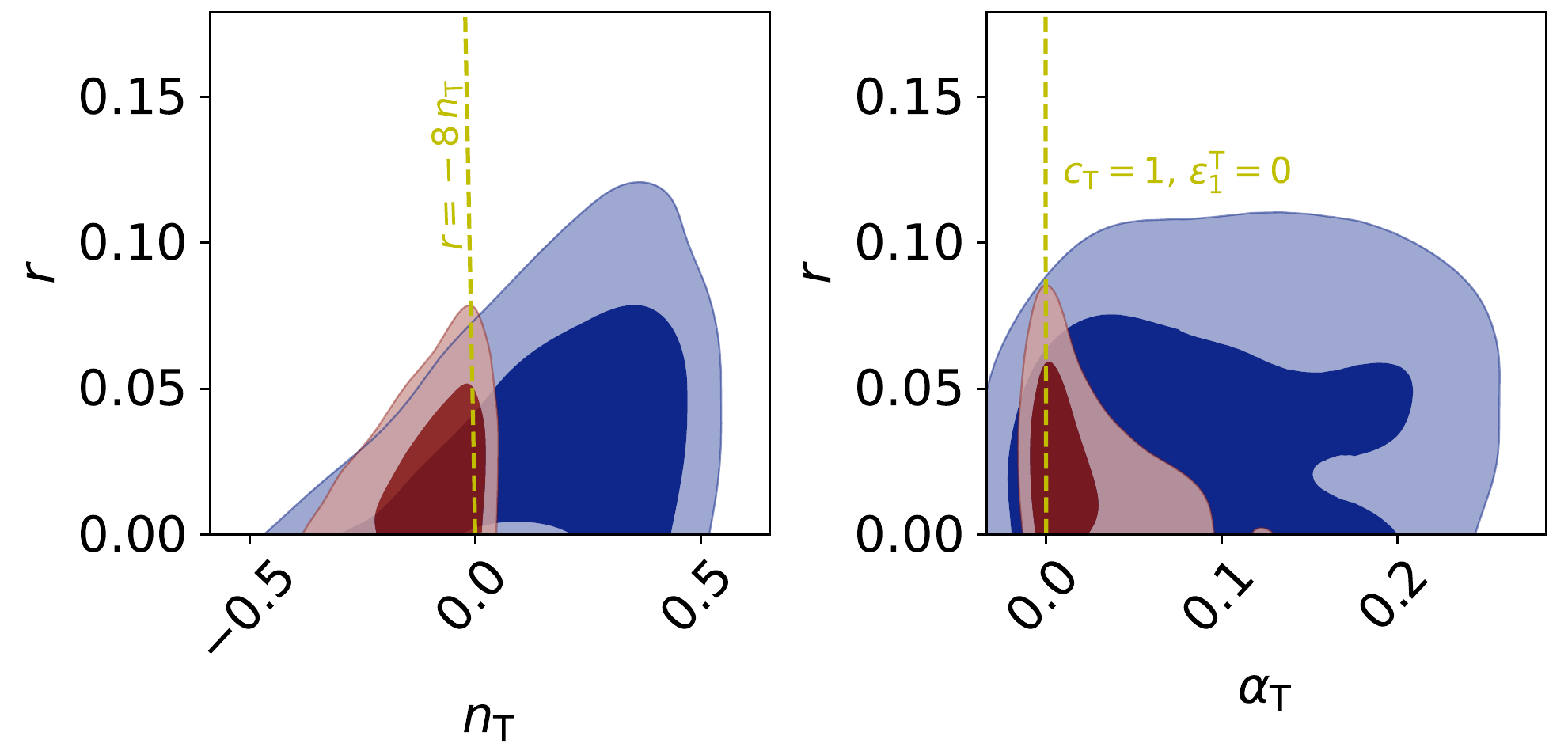}
    \caption{Marginalized 2D posterior in the planes ($n_{\rm T}\,,\,r$) and ($\alpha_{\rm T}\,,\,r$). The blue contours are derived from the combination of Planck 2018 \cite{Aghanim:2019ame,Aghanim:2018oex} and Biceps/Keck 2015 \cite{Ade:2018gkx} data (see Sec. \ref{sec.CMBconstraints}) while the red contours take into account the LIGO/VIRGO data on the stochastic background \cite{LIGO_SGWB-2017,LIGO_SGWB-2019} (see Sec. \ref{sec.combined}). The yellow dashed lines represent the standard slow roll relations in the GR limit \emph{i.e.} $c_{\rm T}=1$ and $\epsilon^{\rm T}_1=0$.}
    \label{fig:figure6}
\end{figure}

As we discussed in Sec. \ref{sec.CMBconstraints}, the LIGO/VIRGO limit on the stochastic background amplitude reduces significantly the allowed parameter space for $\epsilon_1^T$ (see also Fig. \ref{fig:figure5}). Therefore, it is worth combining this small scale bound \eqref{et_limit} with CMB data. We include the LIGO/VIRGO upper bound as a half-Gaussian prior on $\epsilon_1^T$ and we sample the same parameter space using the same method and the same priors as those considered in Sec. \ref{sec.CMBconstraints}. In Table \ref{Tab.CMBresults} we give the constraints on the parameters from a combination of Planck and Biceps/Keck with the LIGO/VIRGO constraints, while in Fig.\ref{fig:figure2} we report their $68\%$ and $95\%$ C.L. contour plots. As one can see neither the inclusion of the small scale data is enough to derive precise constraints on the primordial tensor speed that we found to be $c_{\rm T}>0.22$ at 95\% C.L.. Nevertheless, a proper parametrization of the small scale behavior of the tensor spectrum allows us to set tight constraints on its time dependence parameter $\epsilon^{\rm T}_1=0.082 ^{+0.047}_{-0.11}$ at 68\% C.L. and consequently on the other inflationary parameters. In particular, we constrain the tensor-to-scalar ratio $r<0.0599$   at 95\% C.L., which is in perfect agreement with the constraints derived by the Planck Collaboration \cite{Akrami:2018odb}. We also constrain the tensor tilt to be $n_{\rm T}=-0.084^{+0.10}_{-0.047}$ at 68\% C.L. and its running $\alpha_{\rm T}=0.0141^{+0.0035}_{-0.022}$ always at 68\% C.L. These constraints show an improvement of more than an order of magnitude with respect to those derived in Sec. \ref{sec.CMBconstraints} only from the Planck and Biceps/Keck data. Moreover using \eqref{Eq.betaT} we can obtain  derived constraints on the second-order running $\beta_{\rm T}$, namely $\beta_{\rm T}=-0.0061^{+0.011}_{-0.0027}$ at 68\% C.L., again one order of magnitude better than our estimation provided in Sec. \ref{sec.CMBconstraints}.
For completeness we report also the constraints on the other slow roll parameters that can be derived according to the consistency relation discussed in Sec. \ref{sec.generalcons}. We obtain the following constraints from the combination P18+BK15+LV:
\begin{align}
    &\epsilon_1 <  0.00276 &(95\%\ \rm{C.L}) \\
    &\epsilon_2 = 0.0347\pm 0.0046 &(68\%\ \rm{C.L}) \\
    &\epsilon_3 = 0.21\pm 0.22  &(68\%\ \rm{C.L}) 
\end{align}  
Our almost model-independent constraints\footnote{We remember that we have only assumed slow roll inflation and a possible slow time variation of the primordial tensor speed.} on the inflationary parameters reduce significantly the parameter space allowed for models of inflation with non-trivial tensor speed. Indeed the positive (negative) values of $n_{\rm T}$ ($\alpha_{\rm T}$) are now very tightly constrained (see also Fig. \ref{fig:figure6}). This means that a future detection of a large positive (negative) tensor tilt (running), allowed by the present bounds once the generalized consistency relations are relaxed, cannot be brought back to a time variation of the primordial tensor speed, as our results proved.
Besides, thanks to the great improvement in the constraints derived combining the CMB and small scales data, one can better test gravity on the inflationary energy scale. We would like to stress that the generalized consistency relations obtained in Sec. \ref{sec.generalcons} and assumed in our MCMC analysis, generalize the standard slow roll relations that we prove to be recovered when the GR prescriptions $c_{\rm T}=1$ and $\epsilon^{\rm T}_1=0$ are restored. As any departure from these prescriptions would imply physics beyond GR on the inflationary energy scales, it is important to check the consistency between the constraints and the standard slow roll predictions in the GR framework. Let us start noting that the condition $\epsilon^{\rm T}_1=0$ that ensures a constant propagating speed $c_{ \rm T}$ is consistent with our constraints within one standard deviation. Moreover in Fig. \ref{fig:figure6} we plot the 2D marginalized contours at 68\% and 95\% C.L. in the planes $(n_{\rm T}\,,\,r)$ and $(\alpha_{\rm T}\,,\,r)$. The standard consistency relations, yellow dashed lines in the figure, are consistent with our constraints and, above all when the small scale limit \eqref{LVlimit} is included, no significant deviations are observed.

We can, therefore, conclude that our results, even not strong enough to definitively exclude departures from GR on the inflationary energy scales, set interesting constraints on the inflationary models with non-trivial tensor speed, significantly reducing the allowed parameter space for such models. Moreover, they show remarkable accordance between the current data and the standard predictions expected in a GR slow roll scenario. In particular only deviations from GR of the order of $few \times 10^{-1}$ are allowed to combine large and small scale data for models with non-trivial tensor speed (see Fig. \ref{fig:figure6})

\section{Conclusion}
\label{sec.conclusion}

In General Relativity the propagating speed of gravitational waves is predicted to be equal to the speed of light and the ground-based interferometers have measured it to be consistent with the GR prediction within a good level of precision on the astrophysical scales \cite{Monitor:2017mdv, Cornish:2017jml,Liu:2020slm}.
Nevertheless, the propagating speed of the so-called primordial gravitational waves (i.e. the tensor modes sourced by the quantum inflationary fluctuations) are still essentially unconstrained \cite{Raveri:2014eea,Creminelli:2014wna,Giovannini:2015kfa,Cai:2015ipa,Cai:2015yza,Cai:2016ldn,Fumagalli:2016afy}. Albeit a direct detection of PGWs is still missing, the CMB data constrain their amplitude to be much smaller with respect to the primordial scalar perturbations and constraints on the inflationary parameters can be derived \cite{Akrami:2018odb}. Moreover, also small scales experiments on Gravitational Waves, being sensitive to the stochastic background \cite{Akrami:2018odb,LIGO_SGWB-2017,LIGO_SGWB-2019}, can be used together with the CMB data to improve these constraints. 
As any deviation from a constant $c_{\rm T}=c$ would imply physics beyond GR, constraining the propagating speed of PGWs and its time dependence means to test gravity literally at the earliest moments when the inflation takes place.
Using an effective field theory approach we, therefore, introduced a time-dependent primordial propagating speed $c_{\rm T}(t)$ during inflation, studying its impact on the inflationary parameters. In Sec. \ref{sec.generalcons}, under the assumption of slow-roll inflation, we derived a set of equations that relate the propagating speed $c_{\rm T}(t)$ and its time dependence to the inflationary parameters. These relations generalize the usual consistency relations that are recovered when the GR prescription $c_{\rm T}=c$ is restored. Imposing the above mentioned generalized consistency relations we derived some model-independent constraints on the inflationary parameters. In particular, we performed an MCMC analysis to compare current data with our theoretical model. In Sec. \ref{sec.CMBconstraints} we first derived some constraints from the Planck's 2018 temperature and polarization likelihood (which also includes the low multipoles data $\ell < 30$) \cite{Aghanim:2019ame} combined with the lensing likelihood of Planck's 2018 data release based on temperature and polarization lensing reconstruction \cite{Aghanim:2018oex} and the CMB power spectrum likelihood of Bicep2/Keck Array (BK15) \cite{Ade:2018gkx}. We report the results obtained from our MCMC sampling in Table \ref{Tab.CMBresults}. The CMB data alone are not sensitive enough to the primordial propagating speed $c_{\rm T}$ to set a stringent constraint, but they put an upper bound on its time variation per Hubble time $\epsilon^{\rm T}_1<0.203$ at 95\% C.L. defined by Eq. \eqref{e1T}. The fact that the CMB data only set an upper bound on $\epsilon^{\rm T}_1$ is translated into the fact that large positive values of the tensor tilt $n_{\rm T}=0.20^{+0.27}_{-0.13}$ at 68\% C.L.  and its runnings $\alpha_{\rm T}=0.087^{+0.049}_{-0.098}$ at 68\% C.L. are allowed. Nevertheless, as we discussed in the introduction, this region of the parameter space can be more tightly constrained at shorter wavelengths as those probed by ground-based interferometers. In Fig. \eqref{fig:figure1} we plot the constraints from small scale experiments in the plane ($r$ , $n_{\rm T}$) derived under the assumption that the power-law expansion holds from the CMB scales up to such small scales. However, due to the huge difference in the scales proved by CMB and GW data, non-linearities may significantly affect the shape of the primordial spectrum possibly breaking the power-law assumption and the higher-order terms (i.e. the tensor runnings) can lead to nonnegligible corrections. This is why in Sec. \ref{sec.LVconstraints} we generalized the tensor spectrum parametrization to Eq. \eqref{Tensor} including the runnings at any order. As our analysis in Sec. \ref{sec.LVconstraints} proved, positive tensor tilt and positive runnings would strongly amplify the Primordial Gravitational Waves production on small scales as those probed by LIGO and VIRGO. Therefore we used the LIGO/VIRGO upper limit on the stochastic background Eq.\eqref{LVlimit} to derive a tight lower bound $\epsilon_1^{\rm T}\gtrsim - 0.025$ that is translated into the upper bound $n_{\rm T}\lesssim 0.025$ for the tensor tilt. We also used the limit on $\epsilon_1^{\rm T}$ to derive a lower bound for the propagating speed $c_{\rm T}$. In fact excluding superluminal velocities and assuming a linear dismissing of the tensor speed for the whole $\Delta N\simeq 60$ e-folds of inflation we find $c_{\rm T}\gtrsim 0.4\,c$. This result is consistent with the 2D marginalized posteriors shown in Fig.\ref{fig:figure2} that at least within the 68\% C.L. contours seem to prefer values of $c_{\rm T}$ larger than 0.4 times the speed of light. As shown in Fig. \ref{fig:figure5}, once the small scale constraints are considered, a large range of the parameter space allowed by the CMB data now becomes excluded. Therefore in Sec. \ref{sec.combined} we decided to combine the constraints on small scales with the constraints from the CMB performing a new MCMC sampling. To include the small scale bounds derived in sec \ref{sec.LVconstraints}, we introduced a Half-Gaussian prior on the parameter $\epsilon_1^{\rm T}$. The results obtained combining the CMB data and the LIGO/VIRGO  data via the half-Gaussian prior on $\epsilon^{\rm T}_1$ are given in Table \ref{Tab.CMBresults}. Even if the inclusion of the small scale data is not enough to derive precise constraints on the primordial tensor speed - that we found to be $c_{\rm T}>0.22$ at 95\% C.L. - we set tight constraints on its time dependence parameter $\epsilon^{\rm T}_1=0.082 ^{+0.047}_{-0.11}$ at 68\% C.L. and consequently on the other inflationary parameters. In particular we constrain the tensor-to-scalar ratio at the pivot scale $k_*=0.05\rm{Mpc}^{-1}$ to be $r<0.0599$   at 95\% C.L., which is in perfect agreement with the result derived by the Planck Collaboration \cite{Akrami:2018odb}. Moreover we constrain the tensor tilt to be $n_{\rm T}=-0.084^{+0.10}_{-0.047}$ at 68\% C.L., its running $\alpha_{\rm T}=0.0141^{+0.0035}_{-0.021}$ at 68\% C.L. and its running of running $\beta_{\rm T}=-0.0061^{+0.011}_{-0.0027}$ at 68\% C.L.. These constraints show an improvement of more than an order of magnitude with respect to those derived in Sec. \ref{sec.CMBconstraints} considering only the Planck and Biceps/Keck data. 
The constraints we derived in this paper on the inflationary parameters reduce significantly the parameter space allowed for models of inflation with non-trivial tensor speed. Being the positive (negative) values of $n_{\rm T}$ ($\alpha_{\rm T}$) very tightly constrained (see also Fig. \ref{fig:figure6}) a future detection of a large positive (negative) tensor tilt (running) - allowed by the present bounds once the generalized consistency relations are relaxed - cannot be brought back to a time variation of the primordial tensor speed.
Moreover, this improvement in constraints derived combining the CMB and small scale data, allows us to better test gravity on the inflationary energy scale: we have checked the consistency between our constraints and the standard slow roll predictions in the GR framework. The GR prescription $\epsilon^{\rm T}_1=0$ that ensures a constant propagating speed $c_{ \rm T}$ is consistent with our results within 1 standard deviation. Moreover, also the standard consistency relations are perfectly consistent with our constraints, above all when the small scale bounds are included (see also Fig. \ref{fig:figure6}). Since no significant deviations from the standard slow roll predictions are observed, we conclude that even if our results cannot exclude departures from GR on the inflationary energy scales, they significantly constrain models with non-trivial primordial tensor speed, showing good accordance with the predictions excepted in a standard (GR) slow roll scenario.
In the upcoming decade, a new generation of CMB experiments (eg. BICEP3 \cite{BICEP3}, CLASS \cite{CLASS} , SPT-3G \cite{SPT-3G}, Advanced ACTPol \cite{ACTPol}, LBIRD \cite{LBIRD} and CMB-S4 \cite{CMB-S4}) is expected to bring the sensitivity to the amplitude of tensor perturbations down to $r \sim 0.01 - 0.001$ improving the current Planck upper limit around an order of magnitude and possibly leading to the first detection of Primordial Gravitational waves. If so, the generalized consistency relations we derived and the analysis we performed in this paper can be used to definitively check the slow roll predictions and to precisely test gravity on the inflationary energy scales.
\acknowledgments 
W.G. is supported by TASP, iniziativa specifica INFN. F.R. acknowledges support from the NWO and the Dutch Ministry of Education, Culture and Science (OCW), and from the D-ITP consortium, a program of the NWO that is funded by the OCW. 

W.G. and F.R. thank Alessandro Melchiorri, Claudia de Rham and Andrew J. Tolley for the useful Collaboration and for the precious suggestions that contributed to the realization of this article.

In this work we made use of the following python packages that are not mentioned in the text : SciPy \cite{2020SciPy-NMeth} for numerical sampling of the statistical distributions involved in our data analysis, GetDist \cite{GetDist} a tool for the analysis of MCMC samples which employs Matplotlib \cite{Matplotlib} for the realization of the plots in the paper and NumPy \cite{NumPy} for numerical linear algebra.

\appendix
\section{Detailed derivation of \boldmath{$\mathcal P_{\rm T}$} with a time dependent
\boldmath{$c_{\rm T}(t)$}}
\label{App.A}
\label{Appendix_A}
For completeness in this appendix we review in more details the computation of the primordial tensor spectrum with a non-trivial time dependent tensor speed $c_{\rm T}$, showing that under the assumptions  $|\epsilon_1^{\rm T}|\ll 1$, the solution of \eqref{Eq_motion} is given by Eq. \eqref{Solutuion}. First of all, keeping in mind that
\begin{align}
&\frac{d\,a(t)}{d\tau}\doteq a(t)\frac{da(t)}{dt}=a^2(t)\,H\\
&\frac{d\,a^2(t)}{d\tau}=2\,a^3(t) \,H^2 +\mathcal O(\epsilon_1)
\end{align}
we see that for $z_T(t)$ defined in Eq.  \eqref{zT} we have
\begin{align}
\frac{d\,z_T(t)}{d\tau}&=\frac{M_p}{2} a(t)\frac{d}{dt}\left[a(t)c_{\rm T}^{-1}\right]\\&=\frac{M_p}{2} a(t)\left[ \dot a(t) c_{\rm T}^{-1}-a(t) \dot c_{\rm T}\,c_{\rm T}^{-2}  \right]\\
&=\frac{M_p}{2} a(t)\left[ a(t) H c_{\rm T}^{-1}-a(t)\, H\,c_{\rm T}^{-1}  \left(\frac{\dot c_{\rm T}} {H\,c_{\rm T}}\right)  \right]\\
&=\frac{M_p}{2} a(t)^2 \,H c_{\rm T}^{-1}\left[1-\epsilon^{\rm T}_1\right]\\
&=\frac{M_p}{2}\frac{d\,a(t)}{d\tau} c_{\rm T}^{-1}\left[1-\epsilon^{\rm T}_1\right]\\
&\simeq\frac{M_p}{2}\frac{d\,a(t)}{d\tau} c_{\rm T}^{-1} 
\end{align}
and 
\begin{align}
\frac{d^2\,z_T(t)}{d\tau^2}&\simeq\frac{M_p}{2}\left[\frac{d^2\,a(t)}{d\tau^2} c_{\rm T}^{-1} +\frac{d\,a(t)}{d\tau} a(t) \frac{d}{dt}c_{\rm T}^{-1}\right]\\
&\simeq\frac{M_p}{2}\left[\frac{d^2\,a(t)}{d\tau^2} c_{\rm T}^{-1} -\epsilon^{\rm T} a^3(t)\,H^2\,c_{\rm T}^{-1}\right]\\
&\simeq\frac{M_p}{2} c_{\rm T}^{-1}\,\frac{d^2\,a(t)}{d\tau^2}\left[1-\frac{\epsilon^{\rm T}_1}{2}\right]\\
&\simeq\frac{M_p}{2} c_{\rm T}^{-1}\,\frac{d^2\,a(t)}{d\tau^2}
\end{align}
Therefore the equation of motion \eqref{Eq_motion} is equivalent to \eqref{Eq_motion_2} unless corrections of order $|\epsilon^{\rm T}_1|\ll1$. Now we want to prove that $u(\tau,\tilde k)$ given by Eq.  \eqref{Solutuion} correctly solves Eq. \eqref{Eq_motion_2}. First of all, remembering that $\tilde k(t)\doteq c_{\rm T}(t)\, k$, it is worth deriving the following relations:
\begin{align}
&\frac{d\,\tilde k}{d\tau}= -\epsilon^{\rm T}_1\frac{\tilde k}{\tau} \label{dktilde} \\
&\frac{d\,(\tilde k\,\tau)}{d\tau}=\tilde k \left(1-\epsilon^{\rm T}_1\right)\simeq \tilde k\\
&\frac{d}{d\tau} \left(\frac{e^{-i\,\tilde k\,\tau}}{\sqrt{2\tilde k}}\right)\simeq\frac{e^{-i\,\tilde k\,\tau}}{\sqrt{2\tilde k}}\left[-i\,\tilde k+\frac{1}{4}\frac{\epsilon^{\rm T}_1}{\tau}\right]
\end{align}
where in \eqref{dktilde} we have used that in the de Sitter spacetime $\tau\simeq -\left(aH\right)^{-1}$. Now we take the following derivatives:
\begin{align}
\frac{d u}{d\tau}&=\frac{e^{-i\,\tilde k\,\tau}}{\sqrt{2\tilde k}}\left[-i\,\tilde k -\frac{1}{\tau}\left(1-\frac{1}{4}\epsilon^{\rm T}_1\right)+\frac{i}{\tilde k\tau^2}\left(1-\frac{1}{4}\epsilon^{\rm T}_1\right)\right]\\
&\simeq \frac{e^{-i\,\tilde k\,\tau}}{\sqrt{2\tilde k}}\left[-i\,\tilde k -\frac{1}{\tau}+\frac{i}{\tilde k\tau^2}\right]
\end{align}
and finally
\begin{align}
\frac{d^2\,u}{d\tau^2}&\simeq\frac{e^{-i\,\tilde k\,\tau}}{\sqrt{2\tilde k}}\left[  -\tilde k^2 +\frac{i\tilde k}{\tau}\left(1+\frac{3}{4}\epsilon^{\rm T}_1\right)+\frac{2}{\tau^2}\left(1-\frac{1}{8}\epsilon^{\rm T}_1\right)-\frac{2\,i}{\tilde k \tau^3}\left(1-\frac{1}{8}\epsilon^{\rm T}_1\right) \right]\\
&\simeq\frac{e^{-i\,\tilde k\,\tau}}{\sqrt{2\tilde k}}\left[  -\tilde k^2 +\frac{i\tilde k}{\tau}+\frac{2}{\tau^2}-\frac{2\,i}{\tilde k \tau^3}\right]\\
&\simeq\underbrace{\frac{e^{-i\,\tilde k\,\tau}}{\sqrt{2\tilde k}}\left(1-\frac{i}{\tilde k \,\tau}\right)}_{u(\tau,\tilde k)}\left( \frac{2}{\tau^2}-\tilde k ^2\right)\\
&\simeq -\left(\tilde k^2 -\frac{2}{\tau^2}\right)\,u(\tau,\tilde k)
\end{align}
that is noting other than Eq. \eqref{Eq_motion_2}. Therefore, now that we have proved that \eqref{Solutuion} is the correct solution, the derivation of the primordial spectra is trivial: it is sufficient to follow the standard procedure (see e.g. \cite{Riotto:2018pcx, Baumann:2014nda}) with $k\to \tilde k$ that leads us to \eqref{Pt}.

\section{Beyond the linear order in \boldmath{$c_{\rm T}(t)$}}
\label{App.B}
In this paper we have derived some equations that relate the tensor propagating speed $c_{\rm T}$ to the inflationary parameters under the assumption that the second-order time derivative $\ddot c_{\rm T}\simeq 0$. In other words, expanding the propagating speed $c_{\rm T}(t)$ we have taken into account only the linear term. For completeness, we would like to briefly discuss slightly more complicated scenarios in which we consider also the higher-order terms in the Taylor expansion. 

Let us see what happens including also the quadratic term $\ddot c_{\rm T}$: the relation \eqref{e_2t} is modified as follows 
\begin{equation}
\epsilon^{\rm T}_2\doteq \frac{\dot \epsilon^{\rm T}_1}{H\epsilon^{\rm T}_1}=\epsilon_1-\epsilon^{\rm T}_1+\eta_{\rm T}
\end{equation}
where we have to introduce the new parameter
\begin{equation}
\eta_{\rm T} \doteq \frac{\ddot c_{\rm T}}{H\,\dot c_{\rm T}}.
\end{equation}
Neglecting the third order time derivative  $\dddot c_{\rm T}\simeq 0$ we find
\begin{equation}
\frac{d\,\eta_{\rm T} }{d\log k} = \frac{1}{H}\frac{d}{dt} \left[\frac{\ddot c_{\rm T}}{H\,\dot c_{\rm T}} \right]=\eta_{\rm T}\left(\epsilon_1-\eta_{\rm T}\right)
\end{equation}
and $\alpha_{\rm T}$ and $\beta_{\rm T}$ now will read
\begin{equation}
\alpha_{\rm T}=-2\epsilon_1\epsilon_2-\epsilon^{\rm T}_1\left(\epsilon_1-\epsilon^{\rm T}_1+\eta_{\rm T}\right)
\label{alpha_t_2}
\end{equation}
\begin{equation}
\beta_{\rm T}=-2\epsilon_1\epsilon_2^2-2\epsilon_1\epsilon_2\epsilon_3-\epsilon^{\rm T}_1\left[\left(\epsilon_1-\epsilon^{\rm T}_1+\eta_{\rm T}\right)^2 + \epsilon_1\epsilon_2-\epsilon^{\rm T}_1\left(\epsilon_1-\epsilon^{\rm T}_1+\eta_{\rm T}\right) +\eta_{\rm T}\left(\epsilon_1-\eta_{\rm T}\right)\right]
\label{beta_t_2}
\end{equation}
Note that considering the second order derivative of $c_{\rm T}$ with respect to time provides a correction only to the runnings and not to the spectral tilt that in fact is always given by Eq. \eqref{nt}. Moreover even considering the new term $\eta_{\rm T}$ a set of consistency relations can always be derived. Indeed reversing \eqref{alpha_t_2}
\begin{equation}
\eta_{\rm T}=\left(\epsilon_1-\epsilon^{\rm T}_1 \right)+\frac{\alpha_{\rm T}+2\epsilon_1\epsilon_2}{\epsilon^{\rm T}_1}
\label{eta_T}
\end{equation}
and using the Eqs. \eqref{e1}, \eqref{eT}, \eqref{e2}, \eqref{e3} and \eqref{eta_T}, it is easy to see that Eq. \eqref{beta_t_2} still provides a consistency relation for the propagating speed $c_{\rm T}$ and the inflationary parameters.
However in this case the relation will be cubic in the slow roll parameters and will involve also the scalar running $\alpha_{\rm s}$ and the tensor running of running $\beta_{\rm T}$ that are not involved in the respective quadratic relation in the slow roll parameter \eqref{cT_alpha_T} derived under the linear order expansion of $c_{\rm T}$.

This procedure can be generalized at any order: if we expand $c_{\rm T}(t)$ taking all the terms up to the order $n$ and assuming that $\left(\frac{d}{dt}\right)^{n+1} c_{\rm T}\simeq0$, we can always find a consistency relation between $c_{\rm T}$ and the inflationary parameters. This relation will include the scalar runnings up to $\alpha_{n-1}^{\rm S}$ and the tensor runnings up to $\alpha_{n}^{\rm T}$.

Clearly, to test the time dependence of $c_{\rm T}$ beyond the linear expansion, we need an accuracy that we do not have at present. We conclude that the choice to adopt the simply linear approximation for $c_{\rm T}(t)$ is reasonable because it allows us to test its time dependence without complicating the equations or introducing higher-order parameters that will be difficult to constrain with the current cosmological data. We plan to do this in future work.

\section{Superluminal Propagation}
\label{app.C}
In our MCMC sampling we have restricted our attention to the parameter space $c_{\rm T}<1$ excluding  the superluminal propagation.  One may ask if such an artificial exclusion leads to a biased conclusion and, in general, what happens including superluminal velocities. In this appendix we want to clarify some aspects about superluminal velocities and motivate our decision to impose a prior $c_{\rm T}<1$ in our MCMC sampling.

\begin{figure}[h!]
    \centering
    \includegraphics[width=\textwidth]{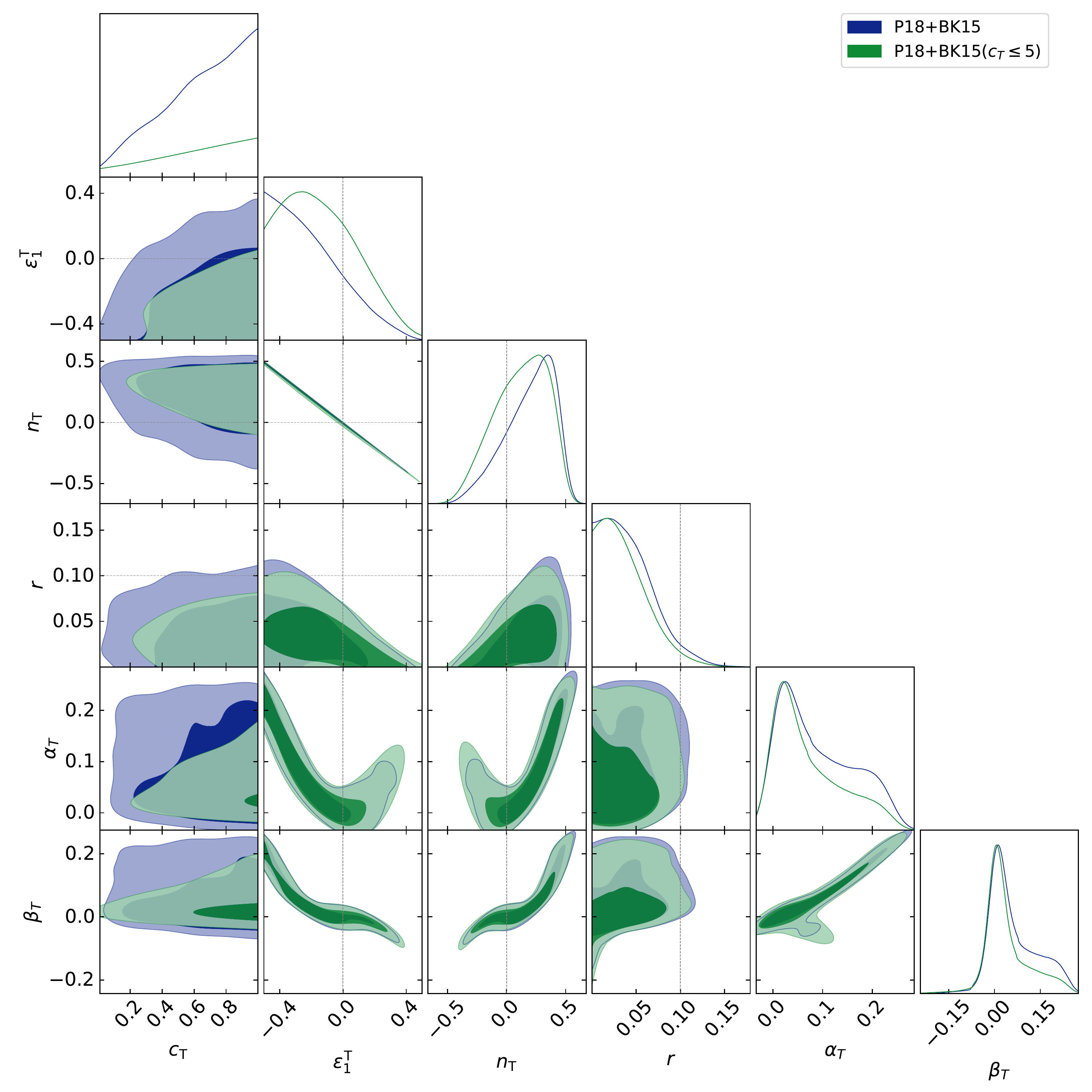}
    \caption{Marginalized 2D and 1D posteriors for the combination of Planck 2018 \cite{Aghanim:2019ame,Aghanim:2018oex} and Biceps/Keck 2015 \cite{Ade:2018gkx} data for the parameters of the tensor spectrum. The blue contours are those obtained exploring only subluminal velocities while the green contours are obtained extending the prior to superluminal velocities $c_{\rm T}<5$. As one can see the choice of exploring only subluminal velocities does not lead to significant bias on the inflationary parameters. Nevertheless, once the superluminal velocities are considered, since the Planck data prefer a vanishing tensor amplitude, the posterior of $c_{\rm T}$ is pushed to $c_{\rm T}\gg 1$ leading to a prior dependent upper (and lower) bound.}
    \label{fig:figure7}
\end{figure}

First of all we want to stress that we have carefully checked that our constraints were not biased by our choice of not exploring superluminal velocities. As a matter of fact, the constraints on $c_{\rm T}$ are almost uncorrelated with the constraints on the other parameters and, even extending our MCMC prior to $c_{\rm T} > 1$, we will end up with almost the same results, see Fig. \ref{fig:figure7}.

We also would like to point out that our theoretical framework holds for both subluminal and superluminal velocities indifferently and that we excluded the superluminal propagation only in our MCMC analysis. This is crucial since from a theoretical point of view imposing subluminal propagation is not as safe an assumption as one may think. In fact, as shown in \cite{deRham:2019ctd, deRham:2020zyh}, depending on the model, it can be possible to perform a change of frame so that in the new frame the tensor speed is $c$, but the speed of the other massless particles is greater than $c $ ending up with a situation where we have actually constrained the speed of normal species to be superluminal, in tension with causality.

However we decided to exclude superluminal velocities from our MCMC analysis for the following reason: as one can see from Eq. \eqref{r},  superluminal velocities will suppress the amplitude of tensor perturbations leading to a completely different phenomenology with respect to subluminal velocities. In fact when $c_T < 1$ the amplitude of the tensor spectrum grows, eventually becoming greater than the Planck experimental error and allowing us to provide a well defined lower bound on the tensor speed. Conversely when $c_T > 1$ the amplitude of the tensor spectrum decreases and the effect of $c_{\rm T}$ on the primordial spectrum is buried in the experimental error, preventing us from achieving a well defined upper bound. 
In other words when the MCMC prior on $c_{\rm T}$ is extended to superluminal velocities,
since the Planck data prefer a vanishing $r$, the posterior distribution of the propagating speed is pushed to $c_{\rm T}\gg1$ and the upper bound on $c_{\rm T}$ is completely dominated by the \textit{a-priori} imposed prior, see Fig. \ref{fig:figure7}.

Furthermore when the prior on the tensor speed is extended to $c_T > 1$ most of the area of the posterior distribution is found for values of $c_{\rm T}$ close to the upper limit of the prior.  Specifically enlarging the prior on $c_{\rm T}$ by a factor of 5 we now get a lower limit $c_T >0.92 c$ ( pushed forward by the same amount with respect to the subluminal case) . 
 
This is clearly a biased result which stems from the fact that we are unable to place an upper bound on the propagating tensor speed with the theoretical framework presented in the paper. The reason behind this is that the MCMC samples accumulate at the higher edge of the imposed range for $c_{\rm T}$ leading to exclude 
values of $c_{\rm T}$ much smaller than the upper limit at more than two standard deviation resulting in a biased lower bound for the tensor speed. Note that this example is merely to show that even pushing the prior on $c_T$ to $c_T \gg 1$ only the posterior of the tensor speed is affected while all other parameters are almost unaffected.

It is also worth noting that to correctly analyze the region $c_T>1$, along with the consistency relation we found, one has to consider also the different phenomenology induced by superluminal propagation. For example a tensor speed different from unity will generate non-gaussian features in the primordial perturbations $f_{\rm NL}\sim 1- c_{\rm T}^2$ \cite{Creminelli:2014wna,Noumi:2014zqa}. Of course this (and other) information can be used to place an upper bound on $c_T>1$, but constraining the superluminal part of $c_{\rm T}$ goes outside the aim of this paper since here we are mainly interested in constraining the shape and amplitude of the tensor spectrum in non-standard theories of inflation with a scale dependent propagating speed. We plan to tackle down the issue of superluminal velocities in a subsequent work.

\section{Extrapolating small scales constraints on $c_{\rm T}$}
\label{app.D}

Even if the main goal of this work was to constrain the shape and the amplitude of the tensor two-point function in a non trivial theory of inflation, in Sec. \ref{sec.combined} we have derived constraints on the propagating speed $c_{\rm T}$ that clearly refer to its value on the CMB scales, with Eqs. \eqref{Pt} and \eqref{Ps} evaluated at the horizon crossing.
In this appendix, we want to discuss the accordance between our results and the current measurement $c_{\rm T}\simeq c$ provided by gravitational experiments. Let us stress that the current observed value $c_{\rm T}\sim c$ refers to the propagating speed of the astrophysical gravitational waves measured by the gravitational detectors on astrophysical scales $k\sim k_{\rm LV}$ and not to the propagating speed of primordial tensor perturbations that are instead generated during the inflationary epoch at energies that can be extremely larger.
We have several observational pieces of evidence that Einstein's theory of general relativity works appropriately on the  astrophysical energy scales, but theoretical arguments suggest that it may need to be modified at high energies and some well motivated extended theories predict a non unitary propagating speed \cite{Horndeski:1974wa,Deffayet:2011gz,Kobayashi:2011nu,Gao:2014soa,Gao:2014fra,Gleyzes:2014qga,Nojiri:2005jg,Makarenko:2017vuk,Bamba:2014zoa,Feng:2020duo,Odintsov:2020xji,Oikonomou:2020sij, Odintsov:2020xji,Odintsov:2020zkl,Satoh:2008ck,Baumann:2015xxa,Oikonomou:2015qha,Haro:2015oqa,Ballesteros:2014sxa,Antoniadis:1993jc,Kawai:1998ab,Soda:1998tr,Kawai:1999pw,Cartier:1999vk,Cartier:2001is,Piao:2003hh}. In our work we have used an effective field theory approach (that, by definition, provides an approximate description of an underlying physical theory at a specific energy scale) to show that if the inflationary energy scale is sufficiently high, high-energy deviations from GR could leave signatures during the inflationary epoch and primordial tensor perturbations could provide a unique observational window to probe gravity at those energy scales.
However, in a consistent theory of gravity, GR has to emerge in the low energies limit in such a way that all the observational evidences for GR (including the observed value $c_{\rm T}\sim c$ on the astrophysical scales) can remain consistent through the evolution of the universe.
Therefore it is worth showing that, our constraints on $c_{\rm T}(k_*)$ are not in conflict with those derived by gravitational detectors.

\begin{figure}[h]
    \centering
    \includegraphics[width=0.6\textwidth]{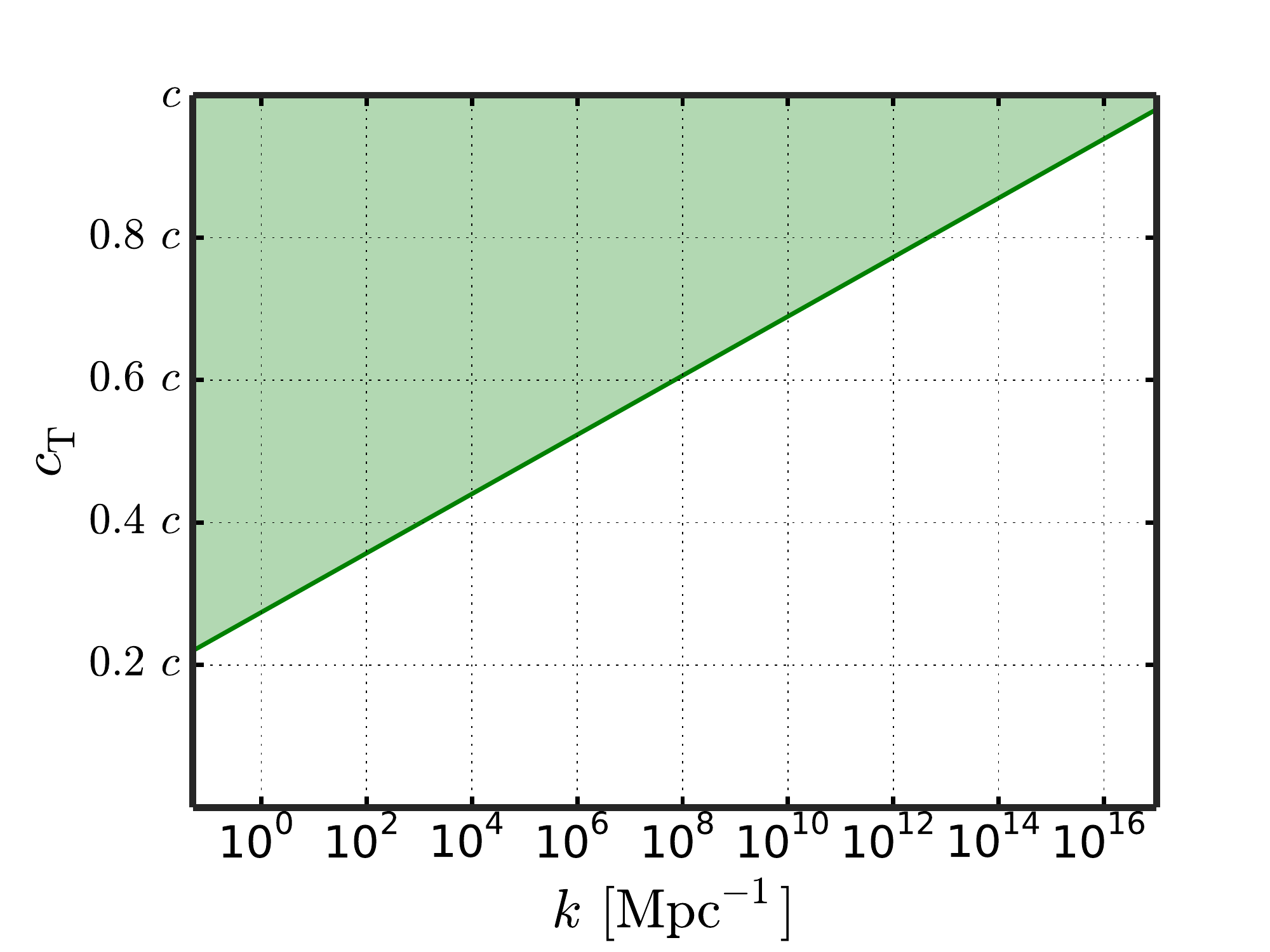}
    \caption{Constraints on the propagating speed $c_{\rm T}$ at the generic scale $k$ extrapolate from the constraints on the CMB scales fixing $\epsilon^{\rm T}_1=0.082$ and $c_{\rm T}(k_*)>0.2$. Remarkably on the LIGO/VIRGO scales we can extrapolate the lower limit $c_{\rm T}(k_{\rm LV})\gtrsim 0.94$, in perfect agreement with the constraints on the astrophysics GWs. }
    \label{fig:figure8}
\end{figure}

Considering the expansion of $\log c_{\rm T}(k)$ we can write
\begin{equation}
    c_{\rm T}(k)=c_{\rm T}(k_*)\left(\frac{k}{k_*}\right)^{\gamma\left(k\right)}
    \label{cT}
\end{equation}
where 
\begin{equation}
    \gamma(k)=\sum_{n=0}^{\infty} \left[\left(\frac{d}{d\log k}\right)^n \epsilon^{\rm T}_1 \right]_{k=k_*} \frac{\log^n\left(\frac{k}{k_*}\right)}{\left(n+1\right)!}
\end{equation}
Because of the discussion provided in sec \ref{sec.LVconstraints}, we can estimate the derivatives of $\epsilon^{\rm T}_1$ as 
\begin{equation}
 \left(\frac{d}{d\log k}\right)^n \epsilon^{\rm T}_1= (-1)^n n! \left(\epsilon^{\rm T}_1\right)^{n+1}
\end{equation}
that gives for $\gamma$
\begin{equation}
    \gamma=-\epsilon^{\rm T}_1\,\frac{\log(1-f(k))}{f(k)}
\end{equation}
where $f(k)=-\epsilon^{\rm T}_1\,\log(k/k_*)$. 
As one can see, the value of the propagating speed at the generic scale $k$ depends on both $c_{\rm T}(k_*)$ and $\epsilon_1^{\rm T}$. Interestingly, using the value derived for $\epsilon^{\rm T}_1\simeq 0.082$, the lower bound for $c_{\rm T}\gtrsim 0.22$ on the CMB scale is translated into the constraints plotted in Fig. \ref{fig:figure8}
at the generic scale $k$.
Even on ultra-high $k$ the power low expansion \eqref{cT} provides reasonable values remarkably close to $c_{\rm T}=c$. In particular on the LIGO/VIRGO scales we have $c_{\rm T}(k_{\rm LV})\gtrsim 0.94\,c$ that is in very good agreement with the constraints on the propagating speed of gravitational waves derived on astrophysical scales \cite{Monitor:2017mdv, Cornish:2017jml,Liu:2020slm}. We therefore conclude that our results are not in conflict with those of gravitational experiments.

\bibliographystyle{aipnum4-1}
\bibliography{main.bib}

\end{document}